\documentclass[superscriptaddress,twocolumn,prb,showpacs]{revtex4-1}

\usepackage{graphicx}
\usepackage{subfigure}

\usepackage{amsmath}
\usepackage{amssymb}
\usepackage{amsthm}
\usepackage{hyperref}

\begin{document}
\title[]{Lattice energies of molecular solids from the random phase approximation 
with singles corrections}

\author{Ji\v{r}\'{i} Klime\v{s}}
\affiliation{J. Heyrovsk\'{y} Institute of Physical Chemistry, Academy of Sciences of the Czech Republic, 
Dolej\v{s}kova 3, CZ-18223 Prague 8, Czech Republic}
\affiliation{Department of Chemical Physics and Optics, Faculty of Mathematics and Physics, 
Charles University, Ke Karlovu 3, CZ-12116 Prague 2, Czech Republic}

\date{\today }

\begin{abstract} 

We use the random phase approximation (RPA) method with the singles correlation
energy contributions to calculate lattice energies of ten molecular solids.
While RPA gives too weak binding, underestimating the reference data by $13.7$\% 
on average, much improved results are obtained when the singles are included 
at the GW singles excitations (GWSE) level, with average absolute difference 
to the reference data of only $3.7$\%.
Consistently with previous results, we find a very good agreement with the reference data
for hydrogen bonded systems, while the binding is too weak for systems where 
dispersion forces dominate.
In fact, the overall accuracy of the RPA+GWSE method is similar to an estimated 
accuracy of the reference data.

\end{abstract}

\maketitle

\section{Introduction}

Molecular solids are an important class of materials both in nature 
and in industries.
They can often have a rich phase diagram or exhibit polymorphism.
The energy differences between different phases or polymorphs can be
very small, on the level of one per cent of the lattice energy.
As the molecules are bound by non-covalent interactions, such minute
differences represent a severe test case for theoretical methods.
For example, density functional theory (DFT) functionals need to be 
augmented with dispersion corrections to, sometimes even qualitatively, describe
systems where there is a competition between dispersion bonding and hydrogen bonds.
While such dispersion corrected DFT (DFT-D) schemes have been successfully applied 
to predict structures of molecular crystals,\cite{bardwell2011} their absolute 
accuracy is still not satisfactory.\cite{otero2012,reilly2013jcp}
One problematic example is the description of the energy differences between 
water ice phases, where many functionals give too large energy 
difference between the common ice I$_{\rm h}$ phase and a high pressure 
ice VIII phase.\cite{santra2011,santra2013,brandenburg2015,bucko2016}

In principle, highly accurate lattice energies can be obtained from methods
based on perturbation theory, in particular using the coupled cluster approach.
This scheme is widely used for studies of molecular clusters where it often
serves to provide benchmark quality data.\cite{jurecka2006}
The coupled cluster method has been applied to obtain lattice energies of 
molecular solids as well, usually to correct data calculated with a simpler
approach.\cite{bludsky2008,beran2010jpcl,wen2011jctc,muller2013,gilliard2014} 
Notably, Yang and coworkers recently obtained very tightly converged lattice
energy of benzene crystal.\cite{yang2014bz}
Because of the compute cost of coupled cluster, the so-called fragment approach
was employed in the majority of these studies.
Here the total lattice energy of the solid is obtained as a sum of interactions
between dimers, trimers, \dots\ of molecules.
To make the scheme computationally feasible, only contributions that are above some 
threshold are considered.
However, it is very challenging to converge the fragment approach fully with respect
to the number of molecules used in the fragments, basis set size, and the order
of perturbation theory.\cite{yang2014bz}
Therefore, there have been several ways proposed that try to improve the convergence, 
such as embedding the fragments in the environment of the 
solid.\cite{bygrave2012,gillan2013a}
Alternatively, implementations of coupled cluster that employ periodic boundary
conditions could be used to obtain lattice energies of molecular solids.
While such schemes are beginning to appear, their large computational cost currently
limits their application to smallest systems only.\cite{booth2013}
Finally, highly accurate lattice energies of molecular solids can be obtained
using quantum Monte Carlo techniques, see Ref.~\onlinecite{dubecky2016} for 
a recent review with summary of the applications to molecular solids and other systems.


There are several methods that don't reach the accuracy of coupled cluster,
but offer much smaller computational cost.
The second order Moller-Plesset perturbation theory (MP2) is the most widely used such scheme
in quantum chemistry and is also available in several codes employing periodic boundary 
conditions.\cite{maschio2007lmp2,marsman2009,grueneis2010,maschio2010,delben2012jctc,delben2013jctc,usvyat2013}
However, the accuracy of MP2 is not satisfactory for systems with delocalized electrons,
for example lattice energies too large, by about 15\%, have been reported for benzene 
and imidazole crystals.\cite{beran2010jpcl}
Several approaches have been proposed to improve the accuracy of MP2 for such systems 
without increasing  
the computational cost substantially.\cite{hesselmann2008,grimme2012,pitonak2010,huang2013MP2C}
Another method that promises a satisfactory accuracy across a range of bonding
situations at a modest computational cost is the random phase approximation (RPA).
RPA often improves the description of systems where simpler semi-local or hybrid
functionals fail, such as for the problem of prediction of the adsorption site of 
CO on metals.\cite{feibelman2001,schimka2010}
Unfortunately, RPA underestimates adsorption energies or the lattice energies
of molecular solids.\cite{li2010rpa,eshuis2012,ren2012,delben2013jctc,klimes2015}
One way to improve the binding energies is to include the so-called renormalized singles corrections
(rSE) to the correlation energy, as proposed by Ren and coworkers.\cite{ren2011,ren2013}
Recently, we have proposed a modification of the singles scheme which includes screening
of the Coulomb interaction and which we call the GW singles excitations (GWSE).\cite{klimes2015}
To gain more understanding about the accuracy of RPA with and without singles corrections 
for the treatment of complex systems, we apply it here to a test set of ten molecular solids.
We find that RPA underbinds molecular solids but the results are dramatically improved
upon adding the singles, especially for hydrogen bonded systems.
In fact, the differences of the RPA+GWSE method from the reference data are 
comparable to the errors and uncertainties present in the reference data
itself.\cite{otero2012,reilly2013jcp}


The application of RPA is not straightforward, especially if one wants 
to converge the values to about 1~kJ/mol or about 1\%, as performed here.
The reason is that the energies strongly depend on the basis set size, the volume of the 
cell used for the reference molecule, and the k-point set used for the solid.
We summarise the convergence behavior here and present a recipe for calculating
converged lattice energies for RPA based methods.

\section{Systems}

We have selected ten molecular crystals from the C21 test set of Otero-de-la-Roza and Johnson
for our study. 
The systems are listed in Table~\ref{tab:mols} together with their
lattice energies, experimental equilibrium volumes, and with the number of molecules
in the unit cell.
We have selected crystals formed by small molecules, such as ammonia, as well as
by comparatively large molecules, such as anthracene with 24 atoms.
Moreover, there are crystals where hydrogen bonding is dominant and also
systems with considerable contribution of dispersion.
This range of system allows us to obtain a valuable information about the performance of the 
RPA based methods in different binding situations.

The initial structures were obtained from the Cambridge Structure Database\cite{allen2002}
and the Crystallography Open Database\cite{grazulis2009} for ammonia\cite{boese1997} 
and carbon dioxide,\cite{simon1980}
the appropriate references to structures are given in Table~\ref{tab:mols}.
The initial structures were geometry optimised using the optB88-vdW 
functional,\cite{dion2004,soler2009,klimes2010,klimes2011} keeping the experimental
unit cell fixed.
The same functional was used to obtain the geometries of isolated molecules, 
for which the initial structures were usually taken from the solid.
The exception is oxalic acid, where the reference molecule contains internal
hydrogen bonds and the molecule was modified appropriately.
To convert from the original .cif files, we have used the OpenBabel program\cite{oboyle2011} 
with our own patch to allow to write files in the POSCAR format required by VASP. 
The patch is included in the development version of OpenBabel.
In some cases VESTA was used to convert the files.\cite{momma2011}
All the structures are available in the supplementary material.\cite{supplementary}

\begin{table}
\caption{Systems selected for the current study, their zero temperature lattice energy 
in kJ/mol as estimated by Reilly and Tkatchenko in Ref.~\onlinecite{reilly2013jcp},
apart from benzene, where the reference value obtained by Yang~{\it et al.} is used,\cite{yang2014bz} 
experimental equilibrium unit cell volume in \AA$^3$, the number of molecules in the unit cell $Z$,
and their CSD\cite{allen2002} code or reference to original article containing 
the structure.}
\label{tab:mols}
\begin{ruledtabular}
\begin{tabular}{lcccc}
System & $E_{\rm latt}$& $V_0$ & $Z$  & CSD code   \\
\hline
Adamantane  & $-69.4$  & 393.07  &  2 & ADAMAN08\\
Anthracene  & $-112.7$  & 456.47  &  2 & ANTCEN09\\
Naphthalene & $-81.7$  & 340.83  &  2 & NAPHTA23\\
Benzene     & $-55.3$  & 481.10  &  4 & BENZEN01\\
CO$_2$      & $-28.4$  & 177.88  &  4 & Ref.~\onlinecite{simon1980}\\
Urea        & $-102.5$  & 145.06  & 2  & UREAXX02\\
Ammonia     & $-37.2$  & 135.05  & 4  & Ref.~\onlinecite{boese1997}\\
Cyanamide   & $-79.7$  & 415.65  & 8  & CYANAM01\\
Oxalic acid $\alpha$ & $-96.3$  & 312.59  & 4 & OXALAC03\\
Oxalic acid $\beta$  & $-96.1$  & 156.87  & 2 & OXALAC04\\
\end{tabular}
\end{ruledtabular}
\end{table}


As we use of the optB88-vdW geometries in the experimental unit cell, our data 
are not the exact RPA lattice energies.
However, since RPA and RPA with singles give rather accurate binding distances,\cite{ren2013} the optimal volume
can be expected to be close to the experimental one.
To estimate the magnitude of the error we used the PBE0-D3$^{\rm BJ}_{\rm ATM}$ 
functional\cite{grimme2010,grimme2011damp,becke2005df,adamo1999}
and compared the energies at the experimental cell to the energies obtained from 
a fit the Murnaghan equation of state (EOS) using seven points around the experimental volume.
For the systems considered here, the energies differ by only $0.4$~kJ/mol on average and 
by $0.7$~kJ/mol at most, see Table~S3 of the SI for details.\cite{supplementary}
Additionally, the optB88-vdW structures for the solid at a given volume and for the molecules 
differ from the optimal RPA structures.
Based on a comparison to structures fully optimised using the PBE-TS functional\cite{tkatchenko2009,bucko2013} 
we estimate the difference in the lattice energy to be usually well below 1~kJ/mol.
Specifically, the lattice energies of fully optimised structures differ from those obtained
from the fit to Murnaghan EOS by $0.4$~kJ/mol on average and by $1.6$~kJ/mol at most
(for oxalic acid $\alpha)$.\cite{subs}
Together with the errors coming from numerical set-up (see below), we estimate 
that our values are within 1--3~kJ/mol of the exact lattice energies of the RPA-based methods.
The reference data have uncertainties as well.
The errors appear in the experimental enthalpies as well as in the corrections that are
used to derive the zero temperature theoretical reference data.
Notably, anharmonic effects are partly accounted for only in the reference
data for adamantane, anthracene, naphthalene, benzene, and urea.
For other systems the reference values might be about 2--3~kJ/mol higher in absolute value.\cite{reilly2013jcp}
While improved accuracy of the reference lattice energies would be greatly valued, 
the current accuracy of the reference is sufficient to draw conclusions about the
performance of the methods in different binding situations.

\section{Computational setup}

We used the VASP program to perform the calculations.\cite{blochl1994,kresse1999}
The RPA correlation energies were obtained using the recently implemented
algorithm with cubic scaling.\cite{kaltak2014rpa2}
The rSE and GWSE energies were calculated
as described in Ref.~\onlinecite{klimes2015}.
The quantities required for the calculations, such as the response function or the self-energy,
are represented on imaginary time or frequency grids, which are obtained using the minimax 
optimisation.\cite{kaltak2014rpa1}
We used 8 frequency and time points to obtain the results.
As suggested in Ref.~\onlinecite{delben2013jctc}, LU-decomposition was used to obtain
the RPA correlation energy, instead of more demanding exact diagonalisation.
The RPA energies were calculated using the usual set-up: the input orbitals and 
energies were obtained from a self-consistent DFT step with the PBE 
exchange-correlation functional.\cite{perdew1996}
All unoccupied states available were included in the calculations to obtain
the RPA, rSE, and GWSE energies, see SI for the details of the set-up.\cite{supplementary}

We used two sets of PAW potentials, standard and hard, 
designated {\tt \_GW } and {\tt \_h\_GW} in the set of PAW potentials distributed with VASP.
We list the details of the potentials in Table~S1 of the SI.\cite{supplementary}
The standard potentials were used to obtain results converged with respect 
to the k-point mesh for solids and the cell size of the reference isolated molecule
as well as with the basis set size.
We then performed calculations with hard potentials at a less dense k-point set and a smaller
unit cell size to correct the converged standard results (see Sec.~{\ref{sec:res:hard}}).
Specifically, a 2$\times$2$\times$2 k-point set for the solid and an 
8$\times$9$\times$10~\AA$^3$ box for the molecule was used, 
apart from anthracene and naphthalene where a 2$\times$2$\times$1 k-point set
and a 7$\times$11$\times$13~\AA$^3$ box was used.
Using finite cell sizes and k-point meshes, the basis set limit of the lattice 
energy was obtained for the hard and normal potentials, giving 
$E^{\rm hard}_{\rm finite}$ and $E^{\rm norm}_{\rm finite}$, respectively. 
The difference $\Delta=E^{\rm hard}_{\rm finite}-E^{\rm norm}_{\rm finite}$ is the desired 
correction. 
The correction was added to $E^{\rm norm}_{\infty}$, which is the lattice energy 
obtained with normal potentials and converged with respect to the numerical parameters,
as follows
\begin{equation}
\label{equ:correct1}
E^{\rm hard}_{\infty}=E^{\rm norm}_{\infty}+( E^{\rm hard}_{\rm finite}-E^{\rm norm}_{\rm finite} )\,
\end{equation}
to obtain the final values of the lattice energy $E^{\rm hard}_{\infty}$
presented here.
We discuss the magnitudes of the corrections in detail in Sec.~\ref{sec:res:hard}.

\begin{table}
\caption{The largest cell side $a$ (in \AA) of the simulation cell used to obtain the RPA correlation
and GWSE energies for the isolated molecule and the largest k-point grid used to
obtain these energies for solid. 
The values are for a plane-wave basis-set cut-off of 600~eV.
Denser grids and larger unit cells were used for smaller cut-offs.}
\label{tab:kpts}
\begin{ruledtabular}
\begin{tabular}{lcccc}
       & \multicolumn{2}{c}{$a$} & \multicolumn{2}{c}{k-pts}  \\
System & RPA & GWSE & RPA & GWSE   \\
\hline
Adamantane  & 12 & 12 & 4$\times$4$\times$4 & 3$\times$3$\times$3 \\
Anthracene  & 12 & 9  & 4$\times$4$\times$4 & 3$\times$3$\times$3  \\
Naphthalene & 11.5&11.5& 3$\times$4$\times$3 & 2$\times$3$\times$2  \\
Benzene     & 12 & 11 & 2$\times$2$\times$2 & 2$\times$2$\times$2  \\
CO$_2$      & 12 & 12 & 4$\times$4$\times$4 & 4$\times$4$\times$4  \\
Urea        & 13 & 11 & 4$\times$4$\times$4 & 4$\times$4$\times$4  \\
Ammonia     & 12 & 12 & 4$\times$4$\times$4 & 4$\times$4$\times$4  \\
Cyanamide   & 11 & 11 & 3$\times$3$\times$2 & 3$\times$3$\times$2  \\
Oxalic acid $\alpha$ & 11& 10 & 3$\times$3$\times$3 & 3$\times$3$\times$3 \\
Oxalic acid $\beta$  & 11& 10 & 3$\times$3$\times$3 & 3$\times$3$\times$3 \\
\end{tabular}
\end{ruledtabular}
\end{table}

The energies of interest are the RPA energy, that is the sum of the 
EXX ($E^{\rm EXX}$) and RPA correlation ($E_{\rm c}^{\rm RPA}$) contributions.
The RPA+GWSE energy $E^{\rm RPA+GWSE}$ is obtained by adding the GWSE correlation 
contribution $E^{\rm GWSE}_{\rm c}$ to the total RPA energy, i.e.
\begin{equation}
E^{\rm RPA+GWSE}=E^{\rm EXX}+E^{\rm RPA}_{\rm c}+E^{\rm GWSE}_{\rm c}\,.
\end{equation}
An equivalent expression is used to obtain the RPA+rSE energy $E^{\rm RPA+rSE}$.

The lattice energy for a given method $M$ is calculated as 
\begin{equation}
\label{eq:coh}
E_{\rm latt}^{\rm M}=E_{\rm sol}^{\rm M}/Z- E_{\rm mol}^{\rm M}\,,
\end{equation}
where $E_{\rm sol}^{\rm M}$ is the energy of the solid, $Z$ is the number
of molecules in the unit cell and $E_{\rm mol}^{\rm M}$ is the energy of the 
isolated molecule.

The energies of the solid and isolated molecule depend on several parameters, such as the
cell volume or the k-point grid, and the cut-off of the plane-wave basis set.
Extrapolations with respect to these parameters need to be performed to obtain converged 
lattice energy.
Our strategy is to use several cut-offs of increasing value and for each of them obtain
the energy of the molecule
converged with the cell volume and the energy of the solid converged with the k-point set.
We use Eq.~\ref{eq:coh} to obtain the lattice energy for every value of cut-off.
The lattice energies are then extrapolated to infinite cut-off (basis set size)
using appropriate convergence behavior, as discussed below.

We consider first the convergence with respect to the volume or the k-point mesh.
The EXX energy converges as $1/V$ with the volume of the cell $V$
for the isolated molecule and as $1/N_k$ with the number of k-points for the solid.
The singles corrections exhibit the same behavior.
Therefore, to obtain converged EXX, rSE, and GWSE energies, we calculated the energies
for increasingly large simulation cells or increasing number of k-points
and extrapolated to infinite volume or k-point set.
For all the molecules apart from naphthalene and anthracene we used boxes of size 
$a\times(a+1)\times(a+2)$~\AA$^3$, starting with $a=7$~\AA.
The boxes for naphthalene and anthracene had dimension $a\times(a+4)\times(a+6)$~\AA$^3$, 
with the smallest $a=6$~\AA. 
The shortest side corresponds to the axis perpendicular to the molecular plane.
The sizes of the boxes were increased in steps of 1~\AA.
For anthracene and naphthalene, steps of 0.5~\AA\ were also used for $a$ larger than 10~\AA.
Compared to the other energies, the calculation of the EXX energy is not computationally demanding
and, apart from extrapolation, we obtained the molecular EXX energies using the screened Coulomb 
potential as well for comparison.\cite{rozzi2006}
These two approaches give almost identical results, differing typically by up to 2~meV.
Unfortunately, the use of the screened Coulomb requires unit cells with sides of about
15~\AA, which are currently not easily accessible for our rSE and GWSE calculations.

For the condensed phases of noble gas solids, Harl and Kresse observed a $1/V^2$ convergence
of the RPA correlation energy with the cell volume.\cite{harl2008}
For molecules, we found this behavior as well.
Therefore, to obtain the RPA correlation energy at infinite cell volume, 
the data were extrapolated assuming the $1/V^2$ convergence.
For small molecules, such as ammonia, urea, carbon dioxide, and oxalic acid, 
the RPA correlation energies are essentially converged for unit cells with the smallest 
side of 10~\AA.
If we found such a behavior, we used the RPA correlation energy calculated at the largest cell.
Finally, for small orbital cut-offs (400~eV and 500~eV) numerical noise can mask the convergence
of the RPA correlation energy and no extrapolation is possible. 
If this situation appeared, we also used the data from the largest cell.
The largest values of $a$ and the most dense k-point grids used to calculate
the RPA correlation and GWSE energies at a plane-wave basis-set cut-off 
of ${\tt ENCUT}=600$~eV are collected in Table~\ref{tab:kpts}.

The memory requirements of the RPA and GWSE calculations grow significantly with the 
basis-set cut-off and with the cell volume used for the isolated molecule or with the number
of k-points used for the solid.
In fact, for large cut-offs (usually above 600~eV), we were not able for some systems 
to acquire all the data that would be required to perform extrapolation to infinite 
cell volume or infinite number of k-points.
To overcome this issue, an estimate of the converged data at large cut-offs 
$E^{\rm large}_{V=\infty}$ was obtained from the converged 
data for small cut-off $E^{\rm small}_{V=\infty}$ and a correction obtained at a smaller 
cell volume as follows:
\begin{equation}
\label{equ:correct}
E^{\rm large}_{V=\infty}=E^{\rm small}_{V=\infty}+( E^{\rm large}_{V=V_0}-E^{\rm small}_{V=V_0} )\,.
\end{equation}
Here, $E^{\rm large}_{V=V_0}$ and $E^{\rm small}_{V=V_0}$
are the energies obtained with the large and small basis set, respectively, and with the cell
volume $V_0$ such that the calculation with large basis-set cut-off was possible.
To control the accuracy of this correction, we calculated it at increasingly large volumes 
until sufficient convergence was achieved.
The same strategy was used to obtain estimates of energies converged with respect 
to the number of k-points. 

We now turn to the dependency of the energies on the cut-offs of the plane-wave basis sets
used in the calculations, that is the orbital basis-set cut-off ({\tt ENCUT} tag in VASP) and 
the cut-off of the basis used to store the response function related properties ({\tt ENCUTGW}
in VASP).
We used several cut-offs, starting from an orbital cut-off of 
400~eV for the normal potentials and a cut-off of 600~eV for the hard potentials.
The cut-off was increased in steps of 100~eV up to at least 800~eV for normal potentials 
and up to at least 900~eV for hard potentials.
The response function cut-off was set to one half of the orbital cut-off throughout.
The EXX and rSE contributions to the lattice energy converge quickly without any clear 
convergence behavior and we used the value at the largest cut-off as the converged number.
Since the EXX energy is less computationally demanding than the other components, 
we increased the cut-off to up to 1000~eV to obtain converged value where necessary.
The change of the EXX lattice energy from the previous lower cut-off is in all cases 1~meV at most.

The RPA correlation and GWSE energies converge usually slowly with the basis set size.
The basis set incompleteness error was found to converge as ${\tt ENCUTGW}^{-3/2}$ in the 
leading order followed by an ${\tt ENCUTGW}^{-5/2}$ term.\cite{harl2008,klimes2014NC,gulans_unp}
If the molecular densities were not overlapping in the solid phase, the leading order would vanish
for the lattice energy and the error would converge as ${\tt ENCUTGW}^{-5/2}$.
We found such convergence only for RPA correlation energies of adamantane.
In all the other cases the convergence rate changes significantly upon forming the solid so that
the lattice energies converge as ${\tt ENCUTGW}^{-3/2}$.
We used the appropriate convergence behavior to extrapolate to infinite basis set size
both for RPA and GWSE.
We note that in a single RPA run VASP calculates the RPA correlation energy 
at several values of {\tt ENCUTGW} and uses this to perform extrapolation to infinite basis set.
We did not rely on these data as they are prone to numerical noise, but we found that for 
large cut-offs the VASP-extrapolated data are close to our extrapolated values.
Specifically, the two approaches agreed to within 0.2~kJ/mol for most of the systems 
when ${\tt ENCUT}=800$~eV was used, the exceptions being adamantane and naphthalene
with differences 0.6~kJ/mol and 1.3~kJ/mol, respectively.
Even for a cut-off of 600~eV, the difference between the RPA correlation energy as extrapolated
within VASP and our extrapolated value is below 1\%\ for all systems, apart from 
adamantane (1.2\%) and naphthalene (1.7\%).
Such accuracy should be acceptable for most of studies.


Overall, for our geometries and hard PAW potentials,
we consider the presented values to be converged to within 0.5~kJ/mol for the small molecules.
For larger molecules, such as anthracene and naphthalene, the uncertainty is larger
and we estimate the error to be approximately 1~kJ/mol.
This comes dominantly from the uncertainties in extrapolating molecular energies 
to infinite volumes.

\section{Results}
\label{sec:results}

\subsection{Lattice energies}
\label{sec:res:coh}

\begin{table}
\caption{Lattice energies of the studied systems in kJ/mol as calculated with
RPA, RPA+rSE, and RPA+GWSE and the reference values derived from 
experimental sublimation enthalpies by Otero-de-la-Roza and Johnson (Ref.~\onlinecite{otero2012}) 
and by Reilly and Tkatchenko (Ref.~\onlinecite{reilly2013jcp}). 
For benzene we use the reference value obtained by Yang and coworkers
instead of the original value presented in Ref.~\onlinecite{reilly2013jcp}.\cite{yang2014bz}
We also show  mean deviations with respect to the latter reference.}
\label{tab:coh}
\begin{ruledtabular}
\begin{tabular}{lccccc}
System & RPA & +rSE & +GWSE & Ref.~\onlinecite{otero2012} & Ref.~\onlinecite{reilly2013jcp} \\
\hline
Adamantane  &$-56.6$   &$-67.1$   &$-67.8$    &$-62.4$   &$-69.4$   \\
Anthracene  &$-92.6$   &$-98.9$   &$-103.5$   &$-100.6$  &$-112.7$\\
Naphthalene  &$-68.4$   &$-73.7$   &$-77.6$    &$-76.3$   &$-81.7$\\
Benzene     &$-45.2$   &$-49.1$   &$-51.5$    &$-50.4$   &$-55.3$\\
CO$_2$      &$-24.1$   &$-26.9$   &$-27.3$    &$-27.8$   &$-28.4$\\
Urea        &$-96.0$   &$-104.7$  &$-104.7$   &$-99.4$   &$-102.5$\\
Ammonia     &$-31.5$   &$-37.9$   &$-37.6$    &$-37.6$   &$-37.2$\\
Cyanamide   &$-71.9$   &$-81.4$   &$-82.0$    &$-79.2$   &$-79.7$\\
Oxalic acid $\alpha$ &$-86.8$ &$-98.6$  &$-98.0$ &$-96.0$ &$-96.3$\\
Oxalic acid $\beta$  &$-87.2$ &$-100.1$ &$-99.0$ &$-95.8$ &$-96.1$\\
\hline
MD          &$9.9$ &$2.1$    &$1.0$  &$  $  &$  $\\
MAD         &$9.9 $ &$4.3$   &$2.9$  &$  $  &$  $\\
MRD         &$-13.7$ &$-2.9$ &$-1.5$ &$  $  &$  $\\
MARD        &$13.7 $ &$5.4$  &$3.7$  &$  $  &$  $\\
\end{tabular}
\end{ruledtabular}
\end{table}

\begin{figure}
    \begin{center}
       \includegraphics[width=8cm,clip=true]{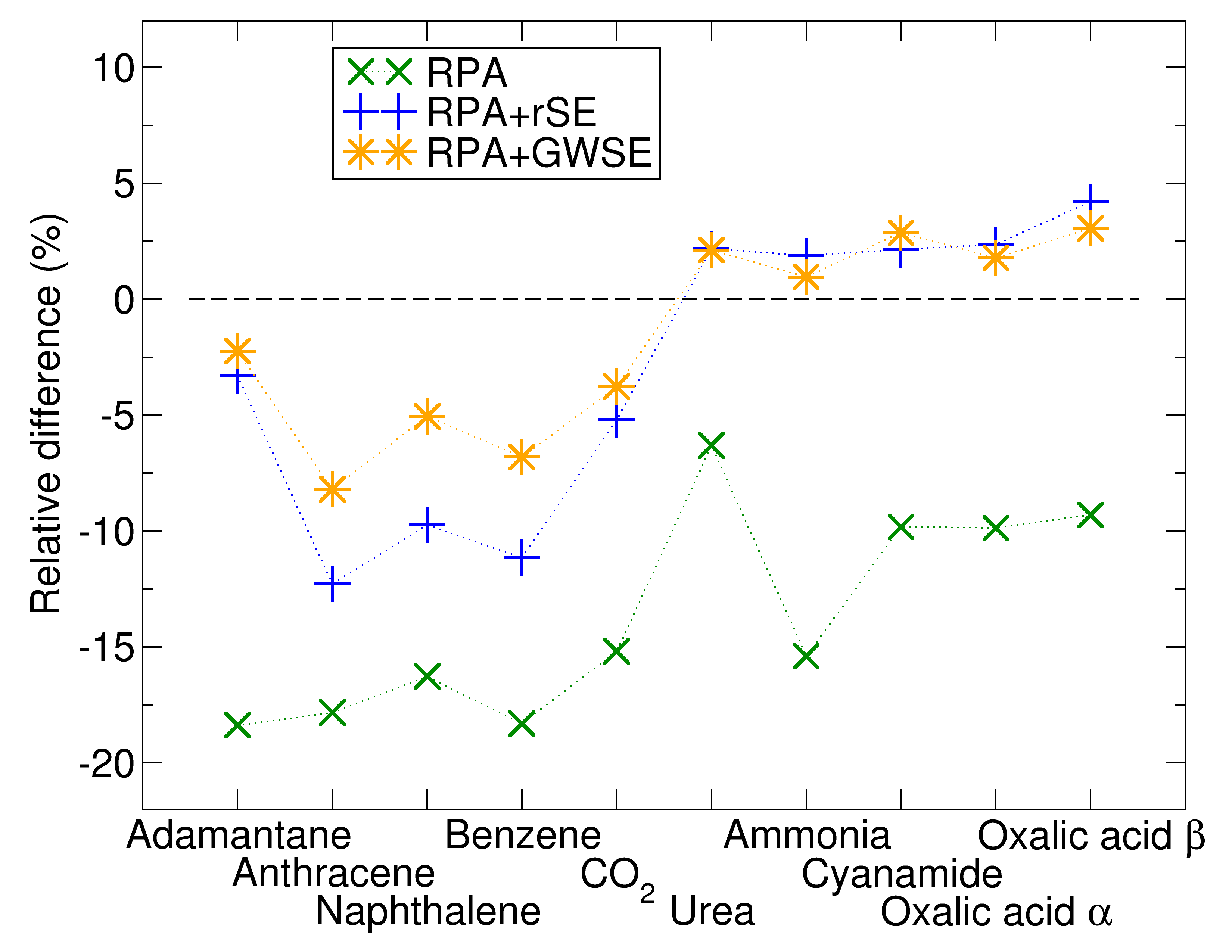}
    \end{center}
   \caption{
Relative deviations of the lattice energies as obtained with RPA based methods
with respect to the reference data of Reilly and Tkatchenko for all systems
apart from benzene where the data of Yang~{\it et al.} is used.\cite{reilly2013jcp,yang2014bz}}
\label{fig:coh}
\end{figure}

The lattice energies calculated with RPA, and RPA with rSE or GWSE contributions
are shown in Table~\ref{tab:coh}. 
We also show the estimates of the reference lattice energies derived from experimental 
sublimation enthalpies by Otero-de-la-Roza and Johnson (Ref.~\onlinecite{otero2012}) 
and by Reilly and Tkatchenko (Ref.~\onlinecite{reilly2013jcp}).
In the latter data, which we use as a reference, we use the value obtained by Yang~{\it et al.}
for the lattice energy of benzene instead of the original number.\cite{yang2014bz} 
The statistics of the differences to the reference data is also given in Table~\ref{tab:coh}.
The relative differences are shown in Fig.~\ref{fig:coh}.
The statistical data are the mean deviation (MD), mean absolute deviation (MAD),
mean relative deviation (MRD), and the mean absolute relative deviation (MARD).
To enable better understanding of the results, we have ordered the systems ascendingly according
to the ratio of the PBE lattice energy and the reference lattice energy.
That is, when going down in the table or from the left to the right in the figure, the relative 
accuracy of PBE increases, corresponding to the change from bonding dominated by dispersion
to bonding dominated by hydrogen bonds.

As one can see, RPA gives lattice energies that underestimate the reference values.
The relative differences are the largest for the dispersion bonded systems, being $-18.4$\% for
adamantane. 
However, as the importance of electrostatics or hydrogen bonding increases, the difference
to the reference is reduced, reaching about $-10$\% for cyanamide and oxalic acid polymorphs.
The relative difference is even smaller for urea, for which we observe a value of $-6.3$\%.
We note that a relative difference of about $-10$\% for hydrogen bonded systems is in agreement 
with the relative errors of lattice energies that we observed previously 
for water ice phases.\cite{macher2014} 
For some of the molecular solids, there are previous RPA calculations of the lattice energies.
Specifically, Galli and coworkers studied the crystals of methane and of benzene, 
finding a lattice energy of $-47$~kJ/mol for the latter, when orbitals from PBE were used.\cite{lu2009,li2010rpa}
This value compares well with our value of $-45.2$~kJ/mol.
More recently, Del Ben and coworkers obtained RPA lattice energies of several molecular
solids.\cite{delben2013jctc}
Their set includes benzene and urea, for which they obtained lattice energies of $-37.6$~kJ/mol and 
$-81.1$~kJ/mol, respectively, to be compared to our values of $-45.2$~kJ/mol and $-96.0$~kJ/mol. 
We believe that the reason for the disagreement is the incomplete convergence with respect to the
supercell size in Ref.~\onlinecite{delben2013jctc}.
For example, a 2$\times$1$\times$2 supercell was used for the benzene crystal, 
so that the shortest side was only 9.435~\AA.


We now turn to the results obtained when the singles corrections are used, 
either at the rSE level (blue crosses in Fig.~\ref{fig:coh}) or at the 
GWSE level (orange stars in Fig.~\ref{fig:coh}).
Adding the singles corrections to RPA data improves the predicted 
lattice energies considerably, e.g., the MARD is reduced from $13.7$\%
of RPA to $5.4$\% for RPA+rSE and to only $3.7$\% for RPA+GWSE.
Both RPA+rSE and RPA+GWSE give excellent agreement with the reference data
for hydrogen bonded systems (urea, ammonia, cyanamide, and oxalic acid polymorphs).
In fact, for these systems the MAD are only 2.2~kJ/mol and 1.9~kJ/mol for
RPA+rSE and RPA+GWSE, respectively.
These values are comparable to the experimental uncertainties for the measurements 
of enthalpies (see the discussion in Ref.~\onlinecite{reilly2013jcp})
and to errors incurred by removing the temperature and quantum nuclear effects.

For systems where dispersion forces dominate, both RPA+rSE and RPA+GWSE improve 
upon RPA, although to a different extent.
Overall, even with singles corrections added, the lattice energies are underestimated
with respect to the reference data.
However, one can observe two distinct situations.
First, for adamantane and CO$_2$, the rSE and GWSE corrections are almost identical.
The results are also rather accurate, with the RPA+GWSE relative differences being
$-2.2$\% for adamantane and $-3.8$\% for CO$_2$.
Second, for aromatic hydrocarbons benzene, naphthalene, and anthracene,
the rSE and GWSE corrections differ significantly.
This is not surprising since GWSE includes screening which is important for aromatic 
hydrocarbons.
For these systems, RPA+GWSE improves the agreement with the reference data 
compared to RPA+rSE, the differences to the reference are reduced by about 5\%.

\subsection{Comparison to dispersion corrected DFT methods}
\label{sec:res:dftd}

\begin{figure}
    \begin{center}
       \includegraphics[width=8cm,clip=true]{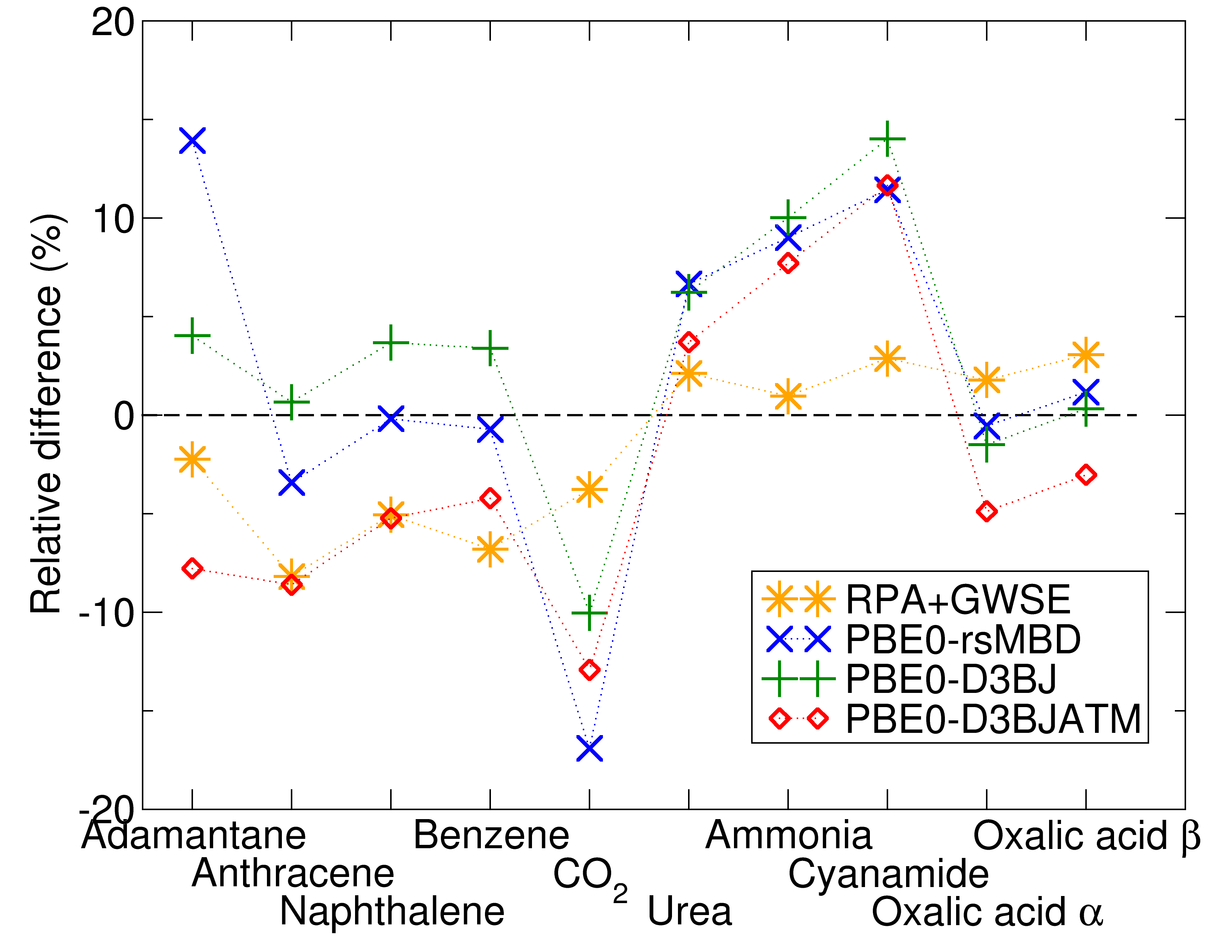}
    \end{center}
   \caption{
Relative deviations of the lattice energies from the reference data as obtained with 
RPA+GWSE and three dispersion corrected DFT functionals.
Reference data of Reilly and Tkatchenko were used for all systems
except for benzene where the data of Yang and coworkers was
used.\cite{reilly2013jcp,yang2014bz}}
\label{fig:cohdft}
\end{figure}

\begin{table}
\caption{Lattice energies of the studied molecular solids in kJ/mol as calculated with
PBE0-rsMBD, PBE0-D3$^{\rm BJ}$, and PBE0-D3$^{\rm BJ}_{\rm ATM}$
and compared to the RPA+GWSE values and the reference data 
of Yang~{\it et al.} for benzene\cite{yang2014bz}
and of Reilly and Tkatchenko (Ref.~\onlinecite{reilly2013jcp}) for the other systems.
We also show  mean deviations with respect to the reference.}
\label{tab:cohdft}
\begin{ruledtabular}
\begin{tabular}{lccccc}
System & PBE0  & PBE0          &PBE0  & RPA & Ref. \\
       & rsMBD & D3$^{\rm BJ}$ &  D3$^{\rm BJ}_{\rm ATM}$& +GWSE & \\
\hline
Adamantane  &$   -79.1 $&$     -72.2  $&$  -64.0$    &$-67.8$     &$-69.4$   \\
Anthracene   &$  -108.8  $&$   -113.4  $&$ -103.0 $ &$-103.5$     &$-112.7$\\
Naphthalene  &$  -81.6   $&$   -84.7  $&$  -77.4 $&$-77.6$      &$-81.7$\\
Benzene       &$ -54.9    $&$  -57.2  $&$ -53.0 $&$-51.5$      &$-55.3$\\
CO$_2$      &$   -23.6   $&$   -25.5  $&$ -24.7 $&$-27.3$      &$-28.4$\\
Urea        &$   -109.3   $&$  -108.9  $&$ -106.3 $ &$-104.7$      &$-102.5$\\
Ammonia      &$  -40.6   $&$   -40.9   $&$ -40.1 $&$-37.6$       &$-37.2$\\
Cyanamide     &$  -88.8   $&$   -90.9   $&$ -89.0 $&$-82.0$       &$-79.7$\\
Oxalic acid $\alpha$ &$-95.8  $&$ -94.9  $&$  -91.6  $  &$-98.0$  &$-96.3$\\
Oxalic acid $\beta$ &$ -97.2  $&$  -96.4   $&$ -93.2 $   &$-99.0$  &$-96.1$\\
\hline
MD          &$-2.0 $ &$-2.6 $  &$ 1.7 $  &$ 1.0 $  &$  $\\
MAD         &$4.0 $ &$3.4 $  &$ 4.9 $  &$2.9  $  &$  $\\
MRD         &$2.0 $ &$3.1 $  &$-2.4  $  &$-1.5  $  &$  $\\
MARD        &$6.4 $ &$5.4 $  &$7.0  $  &$3.7  $  &$  $\\
\end{tabular}
\end{ruledtabular}
\end{table}

It is interesting to compare how do RPA and its modifications compare
to the state-of-the-art DFT functionals.
To this end we obtained the lattice energies of the molecular solids
using the PBE0-D3$^{\rm BJ}$ and 
PBE0-D3$^{\rm BJ}_{\rm ATM}$\cite{grimme2010,grimme2011damp,becke2005df,adamo1999}
and PBE0-rsMBD\cite{ambrosetti2014} functionals, as implemented in 
VASP.\cite{paier2005,bucko2016}
Unlike for RPA, where we need to perform several extrapolations,
the DFT energies can be obtained directly using sufficiently large cells and k-point grids.
Specifically, we used an 18~\AA\ box for the molecules and the k-point grid
corresponded to a supercell with at least 20~\AA\ to a side in each direction.
Moreover, the Coulomb cut-off technique was used to speed-up the convergence
of the Fock exchange term with the cell size and k-point grid ($\tt HFRCUT$ tag in 
VASP).\cite{rozzi2006,spencer2008}
Hard PAW potentials were used with {\tt ENCUT} set to $1000$~eV.
The presented lattice energies were obtained by fitting the energies of seven structures around
the experimental volume to a Murnaghan equation of state.
However, the lattice energies based on the structures at experimental volumes 
differ usually by less than 1~kJ/mol.
More details about the settings and results is given in the SI.\cite{supplementary}

The lattice energies obtained with the dispersion corrected functionals are compared to the 
RPA+GWSE data in Figure~\ref{fig:cohdft} and Table~\ref{tab:cohdft}.
For dispersion bonded systems, the dispersion corrected functionals show similar trends
with the PBE0-D3$^{\rm BJ}_{\rm ATM}$ giving the weakest binding, followed by 
PBE0-rsMBD and by PBE0-D3$^{\rm BJ}$, which gives the strongest binding for most of the systems.
The exceptions from this pattern are adamantane and CO$_2$.
For adamantane, PBE0-rsMBD overestimates the lattice energy by almost 14\%, while 
the relative difference for the -D3 correction (both the standard and the ``ATM" one) 
is similar to the value obtained for anthracene, naphthalene, and benzene.
The CO$_2$ crystal apparently represents a problematic case for 
the dispersion corrected schemes as the lattice energy is underestimated 
by 17\%, 10\%, and 13\% for PBE0-rsMBD,  PBE0-D3$^{\rm BJ}$, 
and PBE0-D3$^{\rm BJ}_{\rm ATM}$, respectively.
It has been argued that higher order contributions would have to be
added to improve the lattice  energy of the CO$_2$ crystal.\cite{reilly2013jcp}
In contrast RPA+GWSE, where no restriction on the order of perturbation theory or 
other is done, underestimates the lattice  energy by 4\%, a value which does not 
deviate from the errors of the other dispersion bonded systems.

For hydrogen bonded systems, the dispersion corrected PBE0 functionals give rather 
inconsistent results: while the lattice  energies of oxalic acid polymorphs are close to
the reference, too large values are obtained for urea, ammonia, and cyanamide crystals,
reaching over 10\% for cyanamide.
This is in contrast to RPA+GWSE, which gives consistently slightly
too large binding.
As with the RPA based methods, we find the ordering of oxalic acid polymorphs
to be reversed compared to the reference and experiment (vide infra).
Overall, considering all the systems, RPA+GWSE gives results that are in a better agreement with the 
reference data, the MAD is lower by 0.5~kJ/mol compared to PBE0-D3$^{\rm BJ}$ and 
by even more compared to PBE0-D3$^{\rm BJ}_{\rm ATM}$ and PBE0-rsMBD.
Also the MARD of 3.7\% for RPA+GWSE is considerably smaller than the values 
obtained for the dispersion 
corrected PBE0 schemes, where values of 6.4\%, 5.4\%, and 7.0\% were obtained for 
PBE0-rsMBD, PBE0-D3$^{\rm BJ}$, and PBE0-D3$^{\rm BJ}_{\rm ATM}$, respectively.

\subsection{Oxalic acid polymorphs}
\label{sec:res:poly}

Some molecular solids exhibit polymorphism, where the molecule can crystallize in
two or more possible structures which are energetically close to each other.
It is both difficult and important to describe the energy differences
and the correct energy ordering.\cite{beran2016}
Therefore, we have included in our test set two low-energy polymorphs of oxalic acid 
-- the $\alpha$ and $\beta$ structures.
The two structures are almost isoenergic, the measured sublimation enthalpy 
of the $\alpha$ structure is larger in magnitude only by 0.07~kJ/mol compared to the $\beta$ 
structure.\cite{otero2012}
Both the estimates of the theoretical zero temperature binding energy, performed in 
Ref.~\onlinecite{otero2012} and Ref.~\onlinecite{reilly2013jcp}
lead to an estimate of the energy difference of about 0.2~kJ/mol.

For all the methods we considered we observe the opposite relation between the energies
of the polymorphs, that is, we find the $\beta$ polymorph to be more stable than the $\alpha$ 
structure.
For example, RPA gives a difference of $-0.4$~kJ/mol and RPA+GWSE even $-1.0$~kJ/mol.
We find similar values, between $-1.1$~kJ/mol and $-1.6$~kJ/mol for the dispersion corrected
PBE0 functionals.
In fact, we tested several dispersion corrected functionals available in VASP and used 
not only our geometries but also the structures of Reilly and Tkatchenko\cite{reilly2013jcp} 
and always found the opposite ordering compared to the experimental data.
Therefore, the reason for the discrepancy is currently not clear and to identify 
it would require reference structures with highly accurate energies.
This would then allow one to make a final judgement about the performance of RPA
and its modifications as well as of other methods. 

Previously, the correct energy ordering of the polymorphs was reported for PBE0-MBD
(note the missing ``rs") in Ref.~\onlinecite{reilly2013jcp}
and for PBE0-D3$^{\rm BJ}$ and PBE0-D3$^{\rm BJ}_{\rm ATM}$ in Ref.~\onlinecite{moellmann2014}.
However, our calculations do not reproduce those results.
As a detailed information about the settings used to perform the calculations 
is not given in Ref.~\onlinecite{moellmann2014}, it's not clear where does the difference occur.
We note that some of the parameters of the calculations can have a substantial 
and a rather unexpected effect on the results.
For example, for the oxalic acid polymorphs, the use of standard PAW potentials increases 
the PBE lattice energies by over 3~kJ/mol, this can be attributed to the oxygen PAW potential
and the presence of short hydrogen bonds.
In contrast, when the hard potentials are used the energy of an oxalic acid dimer taken from 
the $\beta$ crystal structure agrees to within $0.3$~kJ/mol with a reference data obtained 
using all-electron calculation,\cite{g09} see Table~S4 of the SI.\cite{supplementary}
Finally, when the Coulomb cut-off technique is not employed for hybrid functionals, the energies 
strongly depend, at least in VASP, on the k-point set used for the solid or on the cell size
used for the reference molecule.
Consequently, if the energies of the solid and of the molecule are not converged 
an error will occur in the lattice energy.




\subsection{Effect of hard PAW potentials}
\label{sec:res:hard}

\begin{figure}
    \begin{center}
       \includegraphics[width=8cm,clip=true]{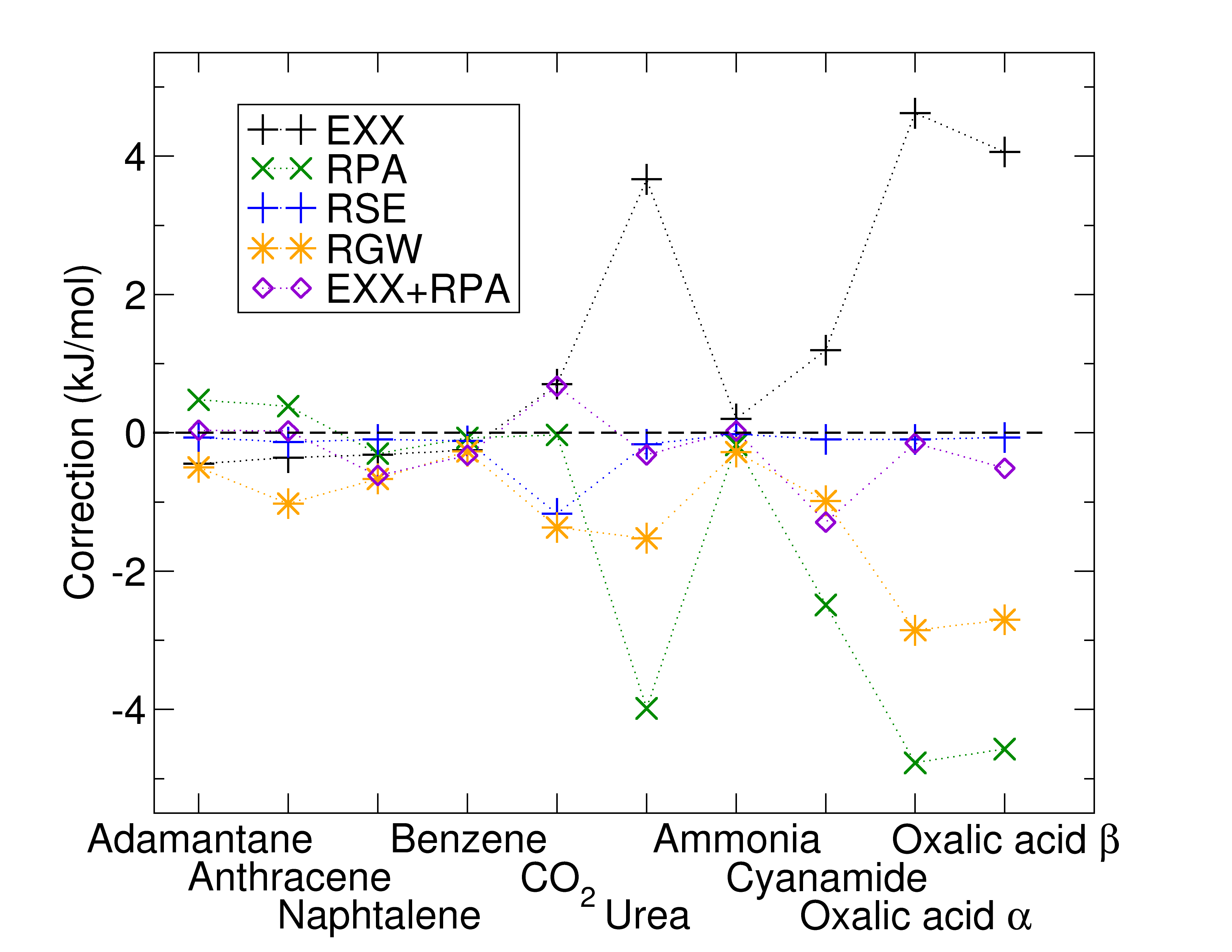}
    \end{center}
   \caption{Differences between lattice energies obtained with hard and normal
PAW potentials.
}
\label{fig:hard}
\end{figure}

We used normal PAW potentials
to converge the results with respect to the number of k-points in the 
solid phase and with the size of the simulation cell for the isolated molecule.
The hard potentials then were used to calculate a correction at a finite cell size
and k-point sampling.
This approach is necessary as the hard potentials require larger basis sets
and consequently have much larger memory and computer time requirements.
It is therefore important to ask if correcting with hard potentials is really required
or if the accuracy of standard potentials is sufficient.

To assess the importance of the corrections with hard potentials, we plot
them in Fig.~\ref{fig:hard} for EXX, RPA, rSE, and GWSE lattice energies.
Moreover, we also show the sum of the EXX and RPA corrections with violet 
diamonds (EXX+RPA). 
The corrections are given in Tables~S5 and~S6 of the SI.\cite{supplementary}
As one can see, the individual EXX and RPA corrections can reach up to 5~kJ/mol in
absolute value for hydrogen bonded systems.
The corrections to rSE are around $-0.1$~kJ/mol for all systems apart from CO$_2$
while the corrections to GWSE can be sizable, increasing the binding by up to 
3~kJ/mol for hydrogen bonded systems.
Importantly, the EXX and RPA corrections tend to cancel
each other so that their sum is below 0.7~kJ/mol in absolute value for most of the systems.
The exception is cyanamide where the total correction reaches $-1.3$~kJ/mol.
Therefore, for many purposes normal PAW potentials are perfectly acceptable for EXX+RPA
and EXX+RPA+rSE calculations.
For systems with short hydrogen bonds, such as oxalic acid polymorphs or urea, 
the corrections to GWSE clearly need to be taken into account if high precision is sought.
However, if that is not the case, normal PAW potentials should be acceptable.

\section{Discussion and Conclusions}

In the present work we calculated the lattice  energies of molecular solids with 
the random phase approximation, either with or without the singles corrections.
The lattice energies obtained by RPA underestimate the reference data by about 13.7\%.
A considerable improvement is found upon including the singles, 
either at the rSE or at the GWSE level.
Specifically, for the RPA+GWSE scheme the mean absolute difference to the reference 
is only 2\% for hydrogen bonded systems while the errors are about twice as large 
for systems where dispersion contribution to the binding dominates.
This is in agreement with the results of Ren and co-workers for the interaction
energies in the S22 test set obtained with RPA+rSE.\cite{ren2013}
Differences to the reference reaching the accuracy of the reference data
have also been found in our previous study for molecular adsorption or for the lattice
constants of atomic solids.\cite{klimes2015}
Together with the cubic scaling with the system size, the good performance makes 
RPA+GWSE a very promising tool for studies of interactions between molecules, 
within solids and for adsorption.

The RPA+GWSE is more computationally and memory demanding compared 
to state-of-the-art dispersion corrected hybrid DFT functionals.
However, it is less prone to outliers, as demonstrated for the CO$_2$ crystal.
Moreover, most of dispersion corrections schemes are targeted at specific systems, 
often organic matter, application to ionic systems or metals can increase the errors
and can require a redesign of the methods.\cite{bucko2013,ruiz2012}
Although more tests need to be done, RPA+GWSE seems to give a good performance across a wide
range of systems with slightly too strong binding for hydrogen bonded systems
and small underbinding for systems where dispersion dominates.
Notably, unlike for the DFT approaches, where there are several parameters one can tune, 
there seems to be no ``easy" way to improve the accuracy of RPA+GWSE.
This is essentially given by the fact that RPA+GWSE is defined only by the terms 
of perturbation theory and by the input PBE orbitals and energies.
One way for a possible improvement is to base RPA on different input orbitals and energies, 
for example the self-consistent ones.\cite{bleiziffer2013,delben2013jctc,klimes2014}
Furthermore, the SOSEX term can be included in the calculations.\cite{grueneis2009}
This is known to improve over RPA for the predictions of atomisation energies 
and related quantities.\cite{paier2012}
The RPA+SOSEX also affects the values of binding energies for weakly bonded systems.
Unfortunately, it was reported that when both singles and SOSEX terms are included, 
the predicted interaction energies for hydrogen bonded systems are worse than if only 
one of them is used.\cite{ren2013}
Finally, one can also go beyond RPA+GWSE by adding an exchange-correlation kernel $f_{xc}$.
Using this approach, improved atomisation energies with little changes of intermolecular 
interaction energies have been recently reported.\cite{olsen2013,olsen2014}

One of the issues we encountered is the accuracy of the reference data.
The differences of the RPA+GWSE data from the reference are similar to estimated
uncertainties in the reference data itself.
For example, the original reference data\cite{otero2012} where a simpler scheme to subtract 
temperature and quantum effects was used has larger statistical deviations from the 
reference data of Reilly of Tkatchenko\cite{reilly2013jcp} than our RPA+GWSE values have.
Moreover, the reference data of Reilly and Tkatchenko for the benzene crystal
differs by  4.6~kJ/mol, {\it i.e.} by about 7\%, from the value obtained 
by the fragment approach and accurate quantum chemical methods by 
Yang and coworkers.\cite{yang2014bz}
Therefore, there is clearly a need for a well defined test set for molecular solids,
with high quality lattice energies and available structures of the molecules and solids.
Producing such data is difficult and computationally demanding, 
but, as demonstrated by the S22 test set for molecular dimers,\cite{jurecka2006} 
will be of a great value. 

\begin{acknowledgments}
JK is supported by the European Union's Horizon 2020 research and innovation
programme under the Marie Sklodowska-Curie grant agreement No 658705.
Computational resources were provided by the MetaCentrum under the program
LM2010005 and the CERIT-SC under the program Centre CERIT Scientific Cloud, 
part of the Operational Program Research and Development for Innovations, 
Reg. no. CZ.1.05/3.2.00/08.0144 and by the IT4Innovations Centre of Excellence 
project (CZ.1.05/1.1.00/02.0070), funded by the European Regional Development Fund 
and the national budget of the Czech Republic via the Research and Development 
for Innovations Operational Programme, as well as Czech Ministry of Education, 
Youth and Sports via the project Large Research, Development and Innovations 
Infrastructures (LM2011033).
We thank J. G. Brandenburg and A. Michaelides for comments on the paper.
\end{acknowledgments}

\section*{References}

\begin{thebibliography}{10}%
\makeatletter
\providecommand \@ifxundefined [1]{%
 \ifx #1\undefined \expandafter \@firstoftwo
 \else \expandafter \@secondoftwo
\fi
}%
\providecommand \@ifnum [1]{%
 \ifnum #1\expandafter \@firstoftwo
 \else \expandafter \@secondoftwo
\fi
}%
\providecommand \enquote [1]{``#1''}%
\providecommand \bibnamefont  [1]{#1}%
\providecommand \bibfnamefont [1]{#1}%
\providecommand \citenamefont [1]{#1}%
\providecommand\href[0]{\@sanitize\@href}%
\providecommand\@href[1]{\endgroup\@@startlink{#1}\endgroup\@@href}%
\providecommand\@@href[1]{#1\@@endlink}%
\providecommand \@sanitize [0]{\begingroup\catcode`\&12\catcode`\#12\relax}%
\@ifxundefined \pdfoutput {\@firstoftwo}{%
 \@ifnum{\z@=\pdfoutput}{\@firstoftwo}{\@secondoftwo}%
}{%
 \providecommand\@@startlink[1]{\leavevmode}%
 \providecommand\@@endlink[0]{}%
}{%
 \providecommand\@@startlink[1]{%
  \leavevmode
  \pdfstartlink
   attr{/Border[0 0 1 ]/H/I/C[0 1 1]}%
   user{/Subtype/Link/A<</Type/Action/S/URI/URI(#1)>>}%
  \relax
 }%
 \providecommand\@@endlink[0]{\pdfendlink}%
}%
\providecommand \url  [0]{\begingroup\@sanitize \@url }%
\providecommand \@url [1]{\endgroup\@href {#1}{\urlprefix}}%
\providecommand \urlprefix [0]{URL }%
\providecommand \Eprint[0]{\href }%
\@ifxundefined \urlstyle {%
  \providecommand \doi [1]{doi:\discretionary{}{}{}#1}%
}{%
  \providecommand \doi [0]{doi:\discretionary{}{}{}\begingroup
  \urlstyle{rm}\Url }%
}%
\providecommand \doibase [0]{http://dx.doi.org/}%
\providecommand \Doi[1]{\href{\doibase#1}}%
\providecommand \bibAnnote [3]{%
  \BibitemShut{#1}%
  \begin{quotation}\noindent
    \textsc{Key:}\ #2\\\textsc{Annotation:}\ #3%
  \end{quotation}%
}%
\providecommand \bibAnnoteFile [2]{%
  \IfFileExists{#2}{\bibAnnote {#1} {#2} {\input{#2}}}{}%
}%
\providecommand \typeout [0]{\immediate \write \m@ne }%
\providecommand \selectlanguage [0]{\@gobble}%
\providecommand \bibinfo [0]{\@secondoftwo}%
\providecommand \bibfield [0]{\@secondoftwo}%
\providecommand \translation [1]{[#1]}%
\providecommand \BibitemOpen[0]{}%
\providecommand \bibitemStop [0]{}%
\providecommand \bibitemNoStop [0]{.\EOS\space}%
\providecommand \EOS [0]{\spacefactor3000\relax}%
\providecommand \BibitemShut [1]{\csname bibitem#1\endcsname}%
\bibitem{bardwell2011}%
  \BibitemOpen
  \bibfield{author}{%
  \bibinfo {author} {\bibfnamefont{D.~A.}\ \bibnamefont{Bardwell}}, \bibinfo
  {author} {\bibfnamefont{C.~S.}\ \bibnamefont{Adjiman}}, \bibinfo {author}
  {\bibfnamefont{Y.~A.}\ \bibnamefont{Arnautova}}, \bibinfo {author}
  {\bibfnamefont{E.}~\bibnamefont{Bartashevich}}, \bibinfo {author}
  {\bibfnamefont{S.~X.~M.}\ \bibnamefont{Boerrigter}}, \bibinfo {author}
  {\bibfnamefont{D.~E.}\ \bibnamefont{Braun}}, \bibinfo {author}
  {\bibfnamefont{A.~J.}\ \bibnamefont{Cruz-Cabeza}}, \bibinfo {author}
  {\bibfnamefont{G.~M.}\ \bibnamefont{Day}}, \bibinfo {author}
  {\bibfnamefont{R.~G.}\ \bibnamefont{Della~Valle}}, \bibinfo {author}
  {\bibfnamefont{G.~R.}\ \bibnamefont{Desiraju}}, \bibinfo {author}
  {\bibfnamefont{B.~P.}\ \bibnamefont{van Eijck}}, \bibinfo {author}
  {\bibfnamefont{J.~C.}\ \bibnamefont{Facelli}}, \bibinfo {author}
  {\bibfnamefont{M.~B.}\ \bibnamefont{Ferraro}}, \bibinfo {author}
  {\bibfnamefont{D.}~\bibnamefont{Grillo}}, \bibinfo {author}
  {\bibfnamefont{M.}~\bibnamefont{Habgood}}, \bibinfo {author}
  {\bibfnamefont{D.~W.~M.}\ \bibnamefont{Hofmann}}, \bibinfo {author}
  {\bibfnamefont{F.}~\bibnamefont{Hofmann}}, \bibinfo {author}
  {\bibfnamefont{K.~V.~J.}\ \bibnamefont{Jose}}, \bibinfo {author}
  {\bibfnamefont{P.~G.}\ \bibnamefont{Karamertzanis}}, \bibinfo {author}
  {\bibfnamefont{A.~V.}\ \bibnamefont{Kazantsev}}, \bibinfo {author}
  {\bibfnamefont{J.}~\bibnamefont{Kendrick}}, \bibinfo {author}
  {\bibfnamefont{L.~N.}\ \bibnamefont{Kuleshova}}, \bibinfo {author}
  {\bibfnamefont{F.~J.~J.}\ \bibnamefont{Leusen}}, \bibinfo {author}
  {\bibfnamefont{A.~V.}\ \bibnamefont{Maleev}}, \bibinfo {author}
  {\bibfnamefont{A.~J.}\ \bibnamefont{Misquitta}}, \bibinfo {author}
  {\bibfnamefont{S.}~\bibnamefont{Mohamed}}, \bibinfo {author}
  {\bibfnamefont{R.~J.}\ \bibnamefont{Needs}}, \bibinfo {author}
  {\bibfnamefont{M.~A.}\ \bibnamefont{Neumann}}, \bibinfo {author}
  {\bibfnamefont{D.}~\bibnamefont{Nikylov}}, \bibinfo {author}
  {\bibfnamefont{A.~M.}\ \bibnamefont{Orendt}}, \bibinfo {author}
  {\bibfnamefont{R.}~\bibnamefont{Pal}}, \bibinfo {author}
  {\bibfnamefont{C.~C.}\ \bibnamefont{Pantelides}}, \bibinfo {author}
  {\bibfnamefont{C.~J.}\ \bibnamefont{Pickard}}, \bibinfo {author}
  {\bibfnamefont{L.~S.}\ \bibnamefont{Price}}, \bibinfo {author}
  {\bibfnamefont{S.~L.}\ \bibnamefont{Price}}, \bibinfo {author}
  {\bibfnamefont{H.~A.}\ \bibnamefont{Scheraga}}, \bibinfo {author}
  {\bibfnamefont{J.}~\bibnamefont{van~de Streek}}, \bibinfo {author}
  {\bibfnamefont{T.~S.}\ \bibnamefont{Thakur}}, \bibinfo {author}
  {\bibfnamefont{S.}~\bibnamefont{Tiwari}}, \bibinfo {author}
  {\bibfnamefont{E.}~\bibnamefont{Venuti}},\ and\ \bibinfo {author}
  {\bibfnamefont{I.~K.}\ \bibnamefont{Zhitkov}},\ }%
  \bibfield{journal}{%
  \bibinfo {journal} {Acta Crystallographica Section B}\ }%
  \textbf{\bibinfo {volume} {67}},\ \bibinfo {pages} {535} (\bibinfo {year}
  {2011})%
  \bibAnnoteFile{NoStop}{bardwell2011}%
\bibitem{otero2012}%
  \BibitemOpen
  \bibfield{author}{%
  \bibinfo {author} {\bibfnamefont{A.}~\bibnamefont{Otero-de-la Roza}}\ and\
  \bibinfo {author} {\bibfnamefont{E.~R.}\ \bibnamefont{Johnson}},\ }%
  \bibfield{journal}{%
  \bibinfo {journal} {J. Chem. Phys.}\ }%
  \textbf{\bibinfo {volume} {137}},\ \bibinfo {pages} {054103} (\bibinfo {year}
  {2012})%
  \bibAnnoteFile{NoStop}{otero2012}%
\bibitem{reilly2013jcp}%
  \BibitemOpen
  \bibfield{author}{%
  \bibinfo {author} {\bibfnamefont{A.~M.}\ \bibnamefont{Reilly}}\ and\ \bibinfo
  {author} {\bibfnamefont{A.}~\bibnamefont{Tkatchenko}},\ }%
  \bibfield{journal}{%
  \bibinfo {journal} {J. Phys. Chem.}\ }%
  \textbf{\bibinfo {volume} {139}},\ \bibinfo {pages} {024705} (\bibinfo {year}
  {2013})%
  \bibAnnoteFile{NoStop}{reilly2013jcp}%
\bibitem{santra2011}%
  \BibitemOpen
  \bibfield{author}{%
  \bibinfo {author} {\bibfnamefont{B.}~\bibnamefont{Santra}}, \bibinfo {author}
  {\bibfnamefont{J.}~\bibnamefont{Klime\v{s}}}, \bibinfo {author}
  {\bibfnamefont{D.}~\bibnamefont{Alf\`e}}, \bibinfo {author}
  {\bibfnamefont{A.}~\bibnamefont{Tkatchenko}}, \bibinfo {author}
  {\bibfnamefont{B.}~\bibnamefont{Slater}}, \bibinfo {author}
  {\bibfnamefont{A.}~\bibnamefont{Michaelides}}, \bibinfo {author}
  {\bibfnamefont{R.}~\bibnamefont{Car}},\ and\ \bibinfo {author}
  {\bibfnamefont{M.}~\bibnamefont{Scheffler}},\ }%
  \bibfield{journal}{%
  \bibinfo {journal} {Phys. Rev. Lett.}\ }%
  \textbf{\bibinfo {volume} {107}},\ \bibinfo {pages} {185701} (\bibinfo {year}
  {2011})%
  \bibAnnoteFile{NoStop}{santra2011}%
\bibitem{santra2013}%
  \BibitemOpen
  \bibfield{author}{%
  \bibinfo {author} {\bibfnamefont{B.}~\bibnamefont{Santra}}, \bibinfo {author}
  {\bibfnamefont{J.}~\bibnamefont{Klime\v{s}}}, \bibinfo {author}
  {\bibfnamefont{A.}~\bibnamefont{Tkatchenko}}, \bibinfo {author}
  {\bibfnamefont{D.}~\bibnamefont{Alf\`{e}}}, \bibinfo {author}
  {\bibfnamefont{B.}~\bibnamefont{Slater}}, \bibinfo {author}
  {\bibfnamefont{A.}~\bibnamefont{Michaelides}}, \bibinfo {author}
  {\bibfnamefont{R.}~\bibnamefont{Car}},\ and\ \bibinfo {author}
  {\bibfnamefont{M.}~\bibnamefont{Scheffler}},\ }%
  \bibfield{journal}{%
  \bibinfo {journal} {J. Chem. Phys.}\ }%
  \textbf{\bibinfo {volume} {139}},\ \bibinfo {pages} {154702} (\bibinfo {year}
  {2013})%
  \bibAnnoteFile{NoStop}{santra2013}%
\bibitem{brandenburg2015}%
  \BibitemOpen
  \bibfield{author}{%
  \bibinfo {author} {\bibfnamefont{J.~G.}\ \bibnamefont{Brandenburg}}, \bibinfo
  {author} {\bibfnamefont{T.}~\bibnamefont{Maas}},\ and\ \bibinfo {author}
  {\bibfnamefont{S.}~\bibnamefont{Grimme}},\ }%
  \bibfield{journal}{%
  \bibinfo {journal} {J. Chem. Phys.}\ }%
  \textbf{\bibinfo {volume} {142}},\ \bibinfo {pages} {124104} (\bibinfo {year}
  {2015})%
  \bibAnnoteFile{NoStop}{brandenburg2015}%
\bibitem{bucko2016}%
  \BibitemOpen
  \bibfield{author}{%
  \bibinfo {author} {\bibfnamefont{T.}~\bibnamefont{Bu\v{c}ko}}, \bibinfo
  {author} {\bibfnamefont{S.}~\bibnamefont{Leb\`{e}gue}}, \bibinfo {author}
  {\bibfnamefont{T.}~\bibnamefont{Gould}},\ and\ \bibinfo {author}
  {\bibfnamefont{J.~G.}\ \bibnamefont{\'{A}ngy\'{a}n}},\ }%
  \bibfield{journal}{%
  \bibinfo {journal} {J. Phys.: Cond. Matt.}\ }%
  \textbf{\bibinfo {volume} {28}},\ \bibinfo {pages} {045201} (\bibinfo {year}
  {2016})%
  \bibAnnoteFile{NoStop}{bucko2016}%
\bibitem{jurecka2006}%
  \BibitemOpen
  \bibfield{author}{%
  \bibinfo {author} {\bibfnamefont{P.}~\bibnamefont{Jure\v{c}ka}}, \bibinfo
  {author} {\bibfnamefont{J.}~\bibnamefont{\v{S}poner}}, \bibinfo {author}
  {\bibfnamefont{J.}~\bibnamefont{\v{C}ern\'y}},\ and\ \bibinfo {author}
  {\bibfnamefont{P.}~\bibnamefont{Hobza}},\ }%
  \bibfield{journal}{%
  \bibinfo {journal} {Phys. Chem. Chem. Phys.}\ }%
  \textbf{\bibinfo {volume} {8}},\ \bibinfo {pages} {1985} (\bibinfo {year}
  {2006})%
  \bibAnnoteFile{NoStop}{jurecka2006}%
\bibitem{bludsky2008}%
  \BibitemOpen
  \bibfield{author}{%
  \bibinfo {author} {\bibfnamefont{O.}~\bibnamefont{Bludsk\'y}}, \bibinfo
  {author} {\bibfnamefont{M.}~\bibnamefont{Rube\v{s}}}, \bibinfo {author}
  {\bibfnamefont{P.}~\bibnamefont{Sold\'an}},\ and\ \bibinfo {author}
  {\bibfnamefont{P.}~\bibnamefont{Nachtigal}},\ }%
  \bibfield{journal}{%
  \bibinfo {journal} {J. Chem. Phys.}\ }%
  \textbf{\bibinfo {volume} {128}},\ \bibinfo {pages} {{114102}} (\bibinfo
  {year} {2008})%
  \bibAnnoteFile{NoStop}{bludsky2008}%
\bibitem{beran2010jpcl}%
  \BibitemOpen
  \bibfield{author}{%
  \bibinfo {author} {\bibfnamefont{G.~J.~O.}\ \bibnamefont{Beran}}\ and\
  \bibinfo {author} {\bibfnamefont{S.}~\bibnamefont{Wen}},\ }%
  \bibfield{journal}{%
  \bibinfo {journal} {J. Phys. Chem. Lett.}\ }%
  \textbf{\bibinfo {volume} {1}},\ \bibinfo {pages} {3480} (\bibinfo {year}
  {2010})%
  \bibAnnoteFile{NoStop}{beran2010jpcl}%
\bibitem{wen2011jctc}%
  \BibitemOpen
  \bibfield{author}{%
  \bibinfo {author} {\bibfnamefont{S.}~\bibnamefont{Wen}}\ and\ \bibinfo
  {author} {\bibfnamefont{G.~J.~O.}\ \bibnamefont{Beran}},\ }%
  \bibfield{journal}{%
  \bibinfo {journal} {J. Chem. Theo. Comput.}\ }%
  \textbf{\bibinfo {volume} {7}},\ \bibinfo {pages} {3733} (\bibinfo {year}
  {2011})%
  \bibAnnoteFile{NoStop}{wen2011jctc}%
\bibitem{muller2013}%
  \BibitemOpen
  \bibfield{author}{%
  \bibinfo {author} {\bibfnamefont{C.}~\bibnamefont{M\"{u}ller}}\ and\ \bibinfo
  {author} {\bibfnamefont{D.}~\bibnamefont{Usvyat}},\ }%
  \bibfield{journal}{%
  \bibinfo {journal} {J. Chem. Theo. Comput.}\ }%
  \textbf{\bibinfo {volume} {9}},\ \bibinfo {pages} {5590} (\bibinfo {year}
  {2013})%
  \bibAnnoteFile{NoStop}{muller2013}%
\bibitem{gilliard2014}%
  \BibitemOpen
  \bibfield{author}{%
  \bibinfo {author} {\bibfnamefont{K.}~\bibnamefont{Gilliard}}, \bibinfo
  {author} {\bibfnamefont{O.}~\bibnamefont{Sode}},\ and\ \bibinfo {author}
  {\bibfnamefont{S.}~\bibnamefont{Hirata}},\ }%
  \bibfield{journal}{%
  \bibinfo {journal} {J. Chem. Phys.}\ }%
  \textbf{\bibinfo {volume} {140}},\ \bibinfo {pages} {174507} (\bibinfo {year}
  {2014})%
  \bibAnnoteFile{NoStop}{gilliard2014}%
\bibitem{yang2014bz}%
  \BibitemOpen
  \bibfield{author}{%
  \bibinfo {author} {\bibfnamefont{J.}~\bibnamefont{Yang}}, \bibinfo {author}
  {\bibfnamefont{W.}~\bibnamefont{Hu}}, \bibinfo {author}
  {\bibfnamefont{D.}~\bibnamefont{Usvyat}}, \bibinfo {author}
  {\bibfnamefont{D.}~\bibnamefont{Matthews}}, \bibinfo {author}
  {\bibfnamefont{M.}~\bibnamefont{Sch\"{u}tz}},\ and\ \bibinfo {author}
  {\bibfnamefont{G.~K.}\ \bibnamefont{Chan}},\ }%
  \bibfield{journal}{%
  \bibinfo {journal} {Science}\ }%
  \textbf{\bibinfo {volume} {345}},\ \bibinfo {pages} {640} (\bibinfo {year}
  {2014})%
  \bibAnnoteFile{NoStop}{yang2014bz}%
\bibitem{bygrave2012}%
  \BibitemOpen
  \bibfield{author}{%
  \bibinfo {author} {\bibfnamefont{P.~J.}\ \bibnamefont{Bygrave}}, \bibinfo
  {author} {\bibfnamefont{N.~L.}\ \bibnamefont{Allan}},\ and\ \bibinfo {author}
  {\bibfnamefont{F.~R.}\ \bibnamefont{Manby}},\ }%
  \bibfield{journal}{%
  \bibinfo {journal} {J. Chem. Phys.}\ }%
  \textbf{\bibinfo {volume} {137}},\ \bibinfo {pages} {164102} (\bibinfo {year}
  {2012})%
  \bibAnnoteFile{NoStop}{bygrave2012}%
\bibitem{gillan2013a}%
  \BibitemOpen
  \bibfield{author}{%
  \bibinfo {author} {\bibfnamefont{M.~J.}\ \bibnamefont{Gillan}}, \bibinfo
  {author} {\bibfnamefont{D.}~\bibnamefont{Alf\`e}}, \bibinfo {author}
  {\bibfnamefont{P.~J.}\ \bibnamefont{Bygrave}}, \bibinfo {author}
  {\bibfnamefont{C.~R.}\ \bibnamefont{Taylor}},\ and\ \bibinfo {author}
  {\bibfnamefont{F.~R.}\ \bibnamefont{Manby}},\ }%
  \bibfield{journal}{%
  \bibinfo {journal} {J. Chem. Phys.}\ }%
  \textbf{\bibinfo {volume} {139}},\ \bibinfo {pages} {114101} (\bibinfo {year}
  {2013})%
  \bibAnnoteFile{NoStop}{gillan2013a}%
\bibitem{booth2013}%
  \BibitemOpen
  \bibfield{author}{%
  \bibinfo {author} {\bibfnamefont{H.~G.}\ \bibnamefont{Booth}}, \bibinfo
  {author} {\bibfnamefont{A.}~\bibnamefont{Gr\"uneis}}, \bibinfo {author}
  {\bibfnamefont{G.}~\bibnamefont{Kresse}},\ and\ \bibinfo {author}
  {\bibfnamefont{A.}~\bibnamefont{Alavi}},\ }%
  \bibfield{journal}{%
  \bibinfo {journal} {Nature}\ }%
  \textbf{\bibinfo {volume} {493}},\ \bibinfo {pages} {{365}} (\bibinfo {year}
  {2013})%
  \bibAnnoteFile{NoStop}{booth2013}%
\bibitem{dubecky2016}%
  \BibitemOpen
  \bibfield{author}{%
  \bibinfo {author} {\bibfnamefont{M.}~\bibnamefont{Dubeck\'{y}}}, \bibinfo
  {author} {\bibfnamefont{L.}~\bibnamefont{Mitas}},\ and\ \bibinfo {author}
  {\bibfnamefont{P.}~\bibnamefont{Jure\v{c}ka}},\ }%
  \bibfield{journal}{%
  \bibinfo {journal} {Chem. Rev.}\ }%
  \textbf{\bibinfo {volume} {116}},\ \bibinfo {pages} {5188} (\bibinfo {year}
  {2016})%
  \bibAnnoteFile{NoStop}{dubecky2016}%
\bibitem{maschio2007lmp2}%
  \BibitemOpen
  \bibfield{author}{%
  \bibinfo {author} {\bibfnamefont{D.}~\bibnamefont{Usvyat}}, \bibinfo {author}
  {\bibfnamefont{L.}~\bibnamefont{Maschio}}, \bibinfo {author}
  {\bibfnamefont{F.~R.}\ \bibnamefont{Manby}}, \bibinfo {author}
  {\bibfnamefont{S.}~\bibnamefont{Casassa}}, \bibinfo {author}
  {\bibfnamefont{M.}~\bibnamefont{Sch\"{u}tz}},\ and\ \bibinfo {author}
  {\bibfnamefont{C.}~\bibnamefont{Pisani}},\ }%
  \bibfield{journal}{%
  \bibinfo {journal} {Phys. Rev. B}\ }%
  \textbf{\bibinfo {volume} {76}},\ \bibinfo {pages} {{076102}} (\bibinfo
  {year} {2007})%
  \bibAnnoteFile{NoStop}{maschio2007lmp2}%
\bibitem{marsman2009}%
  \BibitemOpen
  \bibfield{author}{%
  \bibinfo {author} {\bibfnamefont{M.}~\bibnamefont{Marsman}}, \bibinfo
  {author} {\bibfnamefont{A.}~\bibnamefont{Gr\"uneis}}, \bibinfo {author}
  {\bibfnamefont{J.}~\bibnamefont{Paier}},\ and\ \bibinfo {author}
  {\bibfnamefont{G.}~\bibnamefont{Kresse}},\ }%
  \bibfield{journal}{%
  \bibinfo {journal} {J. Chem. Phys.}\ }%
  \textbf{\bibinfo {volume} {{130}}},\ \bibinfo {pages} {{184103}} (\bibinfo
  {year} {{2009}})%
  \bibAnnoteFile{NoStop}{marsman2009}%
\bibitem{grueneis2010}%
  \BibitemOpen
  \bibfield{author}{%
  \bibinfo {author} {\bibfnamefont{A.}~\bibnamefont{Gr\"uneis}}, \bibinfo
  {author} {\bibfnamefont{M.}~\bibnamefont{Marsman}},\ and\ \bibinfo {author}
  {\bibfnamefont{G.}~\bibnamefont{Kresse}},\ }%
  \bibfield{journal}{%
  \bibinfo {journal} {J. Chem. Phys.}\ }%
  \textbf{\bibinfo {volume} {133}},\ \bibinfo {pages} {{074107}} (\bibinfo
  {year} {2010})%
  \bibAnnoteFile{NoStop}{grueneis2010}%
\bibitem{maschio2010}%
  \BibitemOpen
  \bibfield{author}{%
  \bibinfo {author} {\bibfnamefont{L.}~\bibnamefont{Maschio}}, \bibinfo
  {author} {\bibfnamefont{D.}~\bibnamefont{Usvyat}}, \bibinfo {author}
  {\bibfnamefont{M.}~\bibnamefont{Sch\"{u}tz}},\ and\ \bibinfo {author}
  {\bibfnamefont{B.}~\bibnamefont{Civalleri}},\ }%
  \bibfield{journal}{%
  \bibinfo {journal} {J. Chem. Phys.}\ }%
  \textbf{\bibinfo {volume} {{132}}},\ \bibinfo {pages} {{134706}} (\bibinfo
  {year} {{2010}})%
  \bibAnnoteFile{NoStop}{maschio2010}%
\bibitem{delben2012jctc}%
  \BibitemOpen
  \bibfield{author}{%
  \bibinfo {author} {\bibfnamefont{M.~D.}\ \bibnamefont{Ben}}, \bibinfo
  {author} {\bibfnamefont{J.}~\bibnamefont{Hutter}},\ and\ \bibinfo {author}
  {\bibfnamefont{J.}~\bibnamefont{VandeVondele}},\ }%
  \bibfield{journal}{%
  \bibinfo {journal} {J. Chem. Theo. Comput.}\ }%
  \textbf{\bibinfo {volume} {8}},\ \bibinfo {pages} {4177} (\bibinfo {year}
  {2012})%
  \bibAnnoteFile{NoStop}{delben2012jctc}%
\bibitem{delben2013jctc}%
  \BibitemOpen
  \bibfield{author}{%
  \bibinfo {author} {\bibfnamefont{M.~D.}\ \bibnamefont{Ben}}, \bibinfo
  {author} {\bibfnamefont{J.}~\bibnamefont{Hutter}},\ and\ \bibinfo {author}
  {\bibfnamefont{J.}~\bibnamefont{VandeVondele}},\ }%
  \bibfield{journal}{%
  \bibinfo {journal} {J. Chem. Theo. Comput.}\ }%
  \textbf{\bibinfo {volume} {9}},\ \bibinfo {pages} {2654} (\bibinfo {year}
  {2013})%
  \bibAnnoteFile{NoStop}{delben2013jctc}%
\bibitem{usvyat2013}%
  \BibitemOpen
  \bibfield{author}{%
  \bibinfo {author} {\bibfnamefont{D.}~\bibnamefont{Usvyat}},\ }%
  \bibfield{journal}{%
  \bibinfo {journal} {J. Chem. Phys.}\ }%
  \textbf{\bibinfo {volume} {139}},\ \bibinfo {pages} {194101} (\bibinfo {year}
  {2013})%
  \bibAnnoteFile{NoStop}{usvyat2013}%
\bibitem{hesselmann2008}%
  \BibitemOpen
  \bibfield{author}{%
  \bibinfo {author} {\bibfnamefont{A.}~\bibnamefont{Hesselmann}},\ }%
  \bibfield{journal}{%
  \bibinfo {journal} {J. Chem. Phys.}\ }%
  \textbf{\bibinfo {volume} {128}},\ \bibinfo {pages} {144112} (\bibinfo {year}
  {2008})%
  \bibAnnoteFile{NoStop}{hesselmann2008}%
\bibitem{grimme2012}%
  \BibitemOpen
  \bibfield{author}{%
  \bibinfo {author} {\bibfnamefont{S.}~\bibnamefont{Grimme}}, \bibinfo {author}
  {\bibfnamefont{L.}~\bibnamefont{Goerigk}},\ and\ \bibinfo {author}
  {\bibfnamefont{R.~F.}\ \bibnamefont{Fink}},\ }%
  \bibfield{journal}{%
  \bibinfo {journal} {WIREs Comput. Mol. Sci.}\ }%
  \textbf{\bibinfo {volume} {2}},\ \bibinfo {pages} {886} (\bibinfo {year}
  {2012})%
  \bibAnnoteFile{NoStop}{grimme2012}%
\bibitem{pitonak2010}%
  \BibitemOpen
  \bibfield{author}{%
  \bibinfo {author} {\bibfnamefont{M.}~\bibnamefont{Pito\v{n}\'{a}k}}\ and\
  \bibinfo {author} {\bibfnamefont{A.}~\bibnamefont{He{\ss}elmann}},\ }%
  \bibfield{journal}{%
  \bibinfo {journal} {J. Chem. Theo. Comput.}\ }%
  \textbf{\bibinfo {volume} {6}},\ \bibinfo {pages} {168} (\bibinfo {year}
  {2010})%
  \bibAnnoteFile{NoStop}{pitonak2010}%
\bibitem{huang2013MP2C}%
  \BibitemOpen
  \bibfield{author}{%
  \bibinfo {author} {\bibfnamefont{Y.}~\bibnamefont{Huang}}, \bibinfo {author}
  {\bibfnamefont{Y.}~\bibnamefont{Shao}},\ and\ \bibinfo {author}
  {\bibfnamefont{G.~J.~O.}\ \bibnamefont{Beran}},\ }%
  \bibfield{journal}{%
  \bibinfo {journal} {J. Chem. Phys.}\ }%
  \textbf{\bibinfo {volume} {138}},\ \bibinfo {pages} {224112} (\bibinfo {year}
  {2013})%
  \bibAnnoteFile{NoStop}{huang2013MP2C}%
\bibitem{feibelman2001}%
  \BibitemOpen
  \bibfield{author}{%
  \bibinfo {author} {\bibfnamefont{P.~J.}\ \bibnamefont{Feibelman}}, \bibinfo
  {author} {\bibfnamefont{B.}~\bibnamefont{Hammer}}, \bibinfo {author}
  {\bibfnamefont{J.~K.}\ \bibnamefont{Norskov}}, \bibinfo {author}
  {\bibfnamefont{F.}~\bibnamefont{Wagner}}, \bibinfo {author}
  {\bibfnamefont{M.}~\bibnamefont{Scheffler}}, \bibinfo {author}
  {\bibfnamefont{R.}~\bibnamefont{Stumpf}}, \bibinfo {author}
  {\bibfnamefont{R.}~\bibnamefont{Watwe}},\ and\ \bibinfo {author}
  {\bibfnamefont{J.}~\bibnamefont{Dumesic}},\ }%
  \bibfield{journal}{%
  \bibinfo {journal} {{J. Phys. Chem. B}}\ }%
  \textbf{\bibinfo {volume} {{105}}},\ \bibinfo {pages} {{4018}} (\bibinfo
  {year} {{2001}})%
  \bibAnnoteFile{NoStop}{feibelman2001}%
\bibitem{schimka2010}%
  \BibitemOpen
  \bibfield{author}{%
  \bibinfo {author} {\bibfnamefont{L.}~\bibnamefont{Schimka}}, \bibinfo
  {author} {\bibfnamefont{J.}~\bibnamefont{Harl}}, \bibinfo {author}
  {\bibfnamefont{A.}~\bibnamefont{Stroppa}}, \bibinfo {author}
  {\bibfnamefont{A.}~\bibnamefont{Gr\"uneis}}, \bibinfo {author}
  {\bibfnamefont{M.}~\bibnamefont{Marsman}}, \bibinfo {author}
  {\bibfnamefont{F.}~\bibnamefont{Mittendorfer}},\ and\ \bibinfo {author}
  {\bibfnamefont{G.}~\bibnamefont{Kresse}},\ }%
  \bibfield{journal}{%
  \bibinfo {journal} {Nature Mater.}\ }%
  \textbf{\bibinfo {volume} {9}},\ \bibinfo {pages} {741} (\bibinfo {year}
  {2010})%
  \bibAnnoteFile{NoStop}{schimka2010}%
\bibitem{li2010rpa}%
  \BibitemOpen
  \bibfield{author}{%
  \bibinfo {author} {\bibfnamefont{Y.}~\bibnamefont{Li}}, \bibinfo {author}
  {\bibfnamefont{D.}~\bibnamefont{Lu}}, \bibinfo {author}
  {\bibfnamefont{H.-V.}\ \bibnamefont{Nguyen}},\ and\ \bibinfo {author}
  {\bibfnamefont{G.}~\bibnamefont{Galli}},\ }%
  \bibfield{journal}{%
  \bibinfo {journal} {J Phys. Chem. A}\ }%
  \textbf{\bibinfo {volume} {114}},\ \bibinfo {pages} {1944} (\bibinfo {year}
  {2010})%
  \bibAnnoteFile{NoStop}{li2010rpa}%
\bibitem{eshuis2012}%
  \BibitemOpen
  \bibfield{author}{%
  \bibinfo {author} {\bibfnamefont{H.}~\bibnamefont{Eshuis}}\ and\ \bibinfo
  {author} {\bibfnamefont{F.}~\bibnamefont{Furche}},\ }%
  \bibfield{journal}{%
  \bibinfo {journal} {J. Chem. Phys.}\ }%
  \textbf{\bibinfo {volume} {136}},\ \bibinfo {pages} {084105} (\bibinfo {year}
  {2012})%
  \bibAnnoteFile{NoStop}{eshuis2012}%
\bibitem{ren2012}%
  \BibitemOpen
  \bibfield{author}{%
  \bibinfo {author} {\bibfnamefont{X.}~\bibnamefont{Ren}}, \bibinfo {author}
  {\bibfnamefont{P.}~\bibnamefont{Rinke}}, \bibinfo {author}
  {\bibfnamefont{C.}~\bibnamefont{Joas}},\ and\ \bibinfo {author}
  {\bibfnamefont{M.}~\bibnamefont{Scheffler}},\ }%
  \bibfield{journal}{%
  \bibinfo {journal} {J. Mat. Sci.}\ }%
  \textbf{\bibinfo {volume} {47}},\ \bibinfo {pages} {7447} (\bibinfo {year}
  {2012})%
  \bibAnnoteFile{NoStop}{ren2012}%
\bibitem{klimes2015}%
  \BibitemOpen
  \bibfield{author}{%
  \bibinfo {author} {\bibfnamefont{J.}~\bibnamefont{Klime\v{s}}}, \bibinfo
  {author} {\bibfnamefont{M.}~\bibnamefont{Kaltak}}, \bibinfo {author}
  {\bibfnamefont{E.}~\bibnamefont{Maggio}},\ and\ \bibinfo {author}
  {\bibfnamefont{G.}~\bibnamefont{Kresse}},\ }%
  \bibfield{journal}{%
  \bibinfo {journal} {J. Chem. Phys.}\ }%
  \textbf{\bibinfo {volume} {143}},\ \bibinfo {pages} {{102816}} (\bibinfo
  {year} {2015})%
  \bibAnnoteFile{NoStop}{klimes2015}%
\bibitem{ren2011}%
  \BibitemOpen
  \bibfield{author}{%
  \bibinfo {author} {\bibfnamefont{X.}~\bibnamefont{Ren}}, \bibinfo {author}
  {\bibfnamefont{A.}~\bibnamefont{Tkatchenko}}, \bibinfo {author}
  {\bibfnamefont{P.}~\bibnamefont{Rinke}},\ and\ \bibinfo {author}
  {\bibfnamefont{M.}~\bibnamefont{Scheffler}},\ }%
  \bibfield{journal}{%
  \bibinfo {journal} {Phys. Rev. Lett.}\ }%
  \textbf{\bibinfo {volume} {106}},\ \bibinfo {pages} {153003} (\bibinfo {year}
  {2011})%
  \bibAnnoteFile{NoStop}{ren2011}%
\bibitem{ren2013}%
  \BibitemOpen
  \bibfield{author}{%
  \bibinfo {author} {\bibfnamefont{X.}~\bibnamefont{Ren}}, \bibinfo {author}
  {\bibfnamefont{P.}~\bibnamefont{Rinke}}, \bibinfo {author}
  {\bibfnamefont{G.~E.}\ \bibnamefont{Scuseria}},\ and\ \bibinfo {author}
  {\bibfnamefont{M.}~\bibnamefont{Scheffler}},\ }%
  \bibfield{journal}{%
  \bibinfo {journal} {Phys. Rev. B}\ }%
  \textbf{\bibinfo {volume} {88}},\ \bibinfo {pages} {035120} (\bibinfo {year}
  {2013})%
  \bibAnnoteFile{NoStop}{ren2013}%
\bibitem{allen2002}%
  \BibitemOpen
  \bibfield{author}{%
  \bibinfo {author} {\bibfnamefont{F.~H.}\ \bibnamefont{Allen}},\ }%
  \bibfield{journal}{%
  \bibinfo {journal} {Acta Crystallogr. B}\ }%
  \textbf{\bibinfo {volume} {58}},\ \bibinfo {pages} {380} (\bibinfo {year}
  {2002})%
  \bibAnnoteFile{NoStop}{allen2002}%
\bibitem{grazulis2009}%
  \BibitemOpen
  \bibfield{author}{%
  \bibinfo {author} {\bibfnamefont{S.}~\bibnamefont{Grazulis}}, \bibinfo
  {author} {\bibfnamefont{D.}~\bibnamefont{Chateigner}}, \bibinfo {author}
  {\bibfnamefont{R.~T.}\ \bibnamefont{Downs}}, \bibinfo {author}
  {\bibfnamefont{A.~F.~T.}\ \bibnamefont{Yokochi}}, \bibinfo {author}
  {\bibfnamefont{M.}~\bibnamefont{Quiros}}, \bibinfo {author}
  {\bibfnamefont{L.}~\bibnamefont{Lutterotti}}, \bibinfo {author}
  {\bibfnamefont{E.}~\bibnamefont{Manakova}}, \bibinfo {author}
  {\bibfnamefont{J.}~\bibnamefont{Butkus}}, \bibinfo {author}
  {\bibfnamefont{P.}~\bibnamefont{Moeck}},\ and\ \bibinfo {author}
  {\bibfnamefont{A.}~\bibnamefont{Le~Bail}},\ }%
  \bibfield{journal}{%
  \bibinfo {journal} {J. Appl. Crystallogr.}\ }%
  \textbf{\bibinfo {volume} {42}},\ \bibinfo {pages} {726} (\bibinfo {year}
  {2009})%
  \bibAnnoteFile{NoStop}{grazulis2009}%
\bibitem{boese1997}%
  \BibitemOpen
  \bibfield{author}{%
  \bibinfo {author} {\bibfnamefont{R.}~\bibnamefont{Boese}}, \bibinfo {author}
  {\bibfnamefont{N.}~\bibnamefont{Niederpr\"{u}m}}, \bibinfo {author}
  {\bibfnamefont{D.}~\bibnamefont{Bl\"{a}ser}}, \bibinfo {author}
  {\bibfnamefont{A.}~\bibnamefont{Maulitz}}, \bibinfo {author}
  {\bibfnamefont{M.~Y.}\ \bibnamefont{Antipin}},\ and\ \bibinfo {author}
  {\bibfnamefont{P.~R.}\ \bibnamefont{Mallinson}},\ }%
  \bibfield{journal}{%
  \bibinfo {journal} {J. Phys. Chem. B}\ }%
  \textbf{\bibinfo {volume} {101}},\ \bibinfo {pages} {5794} (\bibinfo {year}
  {1997})%
  \bibAnnoteFile{NoStop}{boese1997}%
\bibitem{simon1980}%
  \BibitemOpen
  \bibfield{author}{%
  \bibinfo {author} {\bibfnamefont{A.}~\bibnamefont{Simon}}\ and\ \bibinfo
  {author} {\bibfnamefont{K.}~\bibnamefont{Peters}},\ }%
  \bibfield{journal}{%
  \bibinfo {journal} {Acta Crystallogr. B}\ }%
  \textbf{\bibinfo {volume} {36}},\ \bibinfo {pages} {2750} (\bibinfo {year}
  {1980})%
  \bibAnnoteFile{NoStop}{simon1980}%
\bibitem{dion2004}%
  \BibitemOpen
  \bibfield{author}{%
  \bibinfo {author} {\bibfnamefont{M.}~\bibnamefont{Dion}}, \bibinfo {author}
  {\bibfnamefont{H.}~\bibnamefont{Rydberg}}, \bibinfo {author}
  {\bibfnamefont{E.}~\bibnamefont{Schr\"oder}}, \bibinfo {author}
  {\bibfnamefont{D.~C.}\ \bibnamefont{Langreth}},\ and\ \bibinfo {author}
  {\bibfnamefont{B.~I.}\ \bibnamefont{Lundqvist}},\ }%
  \bibfield{journal}{%
  \bibinfo {journal} {Phys. Rev. Lett.}\ }%
  \textbf{\bibinfo {volume} {92}},\ \bibinfo {pages} {246401} (\bibinfo {year}
  {2004})%
  \bibAnnoteFile{NoStop}{dion2004}%
\bibitem{soler2009}%
  \BibitemOpen
  \bibfield{author}{%
  \bibinfo {author} {\bibfnamefont{G.}~\bibnamefont{Rom\'{a}n-P\'{e}rez}}\ and\
  \bibinfo {author} {\bibfnamefont{J.~M.}\ \bibnamefont{Soler}},\ }%
  \bibfield{journal}{%
  \bibinfo {journal} {Phys. Rev. Lett.}\ }%
  \textbf{\bibinfo {volume} {103}},\ \bibinfo {pages} {096102} (\bibinfo {year}
  {2009})%
  \bibAnnoteFile{NoStop}{soler2009}%
\bibitem{klimes2010}%
  \BibitemOpen
  \bibfield{author}{%
  \bibinfo {author} {\bibfnamefont{J.}~\bibnamefont{Klime\v{s}}}, \bibinfo
  {author} {\bibfnamefont{D.~R.}\ \bibnamefont{Bowler}},\ and\ \bibinfo
  {author} {\bibfnamefont{A.}~\bibnamefont{Michaelides}},\ }%
  \bibfield{journal}{%
  \bibinfo {journal} {J. Phys.: Cond. Matt.}\ }%
  \textbf{\bibinfo {volume} {22}},\ \bibinfo {pages} {022201} (\bibinfo {year}
  {2010})%
  \bibAnnoteFile{NoStop}{klimes2010}%
\bibitem{klimes2011}%
  \BibitemOpen
  \bibfield{author}{%
  \bibinfo {author} {\bibfnamefont{J.}~\bibnamefont{Klime\v{s}}}, \bibinfo
  {author} {\bibfnamefont{D.~R.}\ \bibnamefont{Bowler}},\ and\ \bibinfo
  {author} {\bibfnamefont{A.}~\bibnamefont{Michaelides}},\ }%
  \bibfield{journal}{%
  \bibinfo {journal} {Phys. Rev. B}\ }%
  \textbf{\bibinfo {volume} {83}},\ \bibinfo {pages} {195131} (\bibinfo {year}
  {2011})%
  \bibAnnoteFile{NoStop}{klimes2011}%
\bibitem{oboyle2011}%
  \BibitemOpen
  \bibfield{author}{%
  \bibinfo {author} {\bibfnamefont{N.~M.}\ \bibnamefont{O'Boyle}}, \bibinfo
  {author} {\bibfnamefont{M.}~\bibnamefont{Banck}}, \bibinfo {author}
  {\bibfnamefont{C.~A.}\ \bibnamefont{James}}, \bibinfo {author}
  {\bibfnamefont{C.}~\bibnamefont{Morley}}, \bibinfo {author}
  {\bibfnamefont{T.}~\bibnamefont{Vandemeersch}},\ and\ \bibinfo {author}
  {\bibfnamefont{G.~R.}\ \bibnamefont{Hutchison}},\ }%
  \bibfield{journal}{%
  \bibinfo {journal} {{J. Cheminf.}}\ }%
  \textbf{\bibinfo {volume} {{3}}},\ \bibinfo {pages} {{33}} (\bibinfo {year}
  {{2011}})%
  \bibAnnoteFile{NoStop}{oboyle2011}%
\bibitem{momma2011}%
  \BibitemOpen
  \bibfield{author}{%
  \bibinfo {author} {\bibfnamefont{K.}~\bibnamefont{Momma}}\ and\ \bibinfo
  {author} {\bibfnamefont{F.}~\bibnamefont{Izumi}},\ }%
  \bibfield{journal}{%
  \bibinfo {journal} {{J. Appl. Cryst.}}\ }%
  \textbf{\bibinfo {volume} {{44}}},\ \bibinfo {pages} {{1272}} (\bibinfo
  {year} {{2011}})%
  \bibAnnoteFile{NoStop}{momma2011}%
\bibitem{supplementary}%
  \BibitemOpen
  \enquote{\bibinfo {title} {{\rm See supplementary material at XXX for details
  about the set-up, the optB88-vdW optimised structures, and additional
  data.}}.}\ %
  \bibAnnoteFile{NoStop}{supplementary}%
\bibitem{grimme2010}%
  \BibitemOpen
  \bibfield{author}{%
  \bibinfo {author} {\bibfnamefont{S.}~\bibnamefont{Grimme}}, \bibinfo {author}
  {\bibfnamefont{J.}~\bibnamefont{Antony}}, \bibinfo {author}
  {\bibfnamefont{S.}~\bibnamefont{Ehrlich}},\ and\ \bibinfo {author}
  {\bibfnamefont{H.}~\bibnamefont{Krieg}},\ }%
  \bibfield{journal}{%
  \bibinfo {journal} {J. Chem. Phys.}\ }%
  \textbf{\bibinfo {volume} {132}},\ \bibinfo {pages} {154104} (\bibinfo {year}
  {2010})%
  \bibAnnoteFile{NoStop}{grimme2010}%
\bibitem{grimme2011damp}%
  \BibitemOpen
  \bibfield{author}{%
  \bibinfo {author} {\bibfnamefont{S.}~\bibnamefont{Grimme}}, \bibinfo {author}
  {\bibfnamefont{S.}~\bibnamefont{Ehrlich}},\ and\ \bibinfo {author}
  {\bibfnamefont{L.}~\bibnamefont{Goerigk}},\ }%
  \bibfield{journal}{%
  \bibinfo {journal} {J. Comput. Chem.}\ }%
  \textbf{\bibinfo {volume} {32}},\ \bibinfo {pages} {1456} (\bibinfo {year}
  {2011})%
  \bibAnnoteFile{NoStop}{grimme2011damp}%
\bibitem{becke2005df}%
  \BibitemOpen
  \bibfield{author}{%
  \bibinfo {author} {\bibfnamefont{A.~D.}\ \bibnamefont{Becke}}\ and\ \bibinfo
  {author} {\bibfnamefont{E.~R.}\ \bibnamefont{Johnson}},\ }%
  \bibfield{journal}{%
  \bibinfo {journal} {J. Chem. Phys.}\ }%
  \textbf{\bibinfo {volume} {123}},\ \bibinfo {pages} {154101} (\bibinfo {year}
  {2005})%
  \bibAnnoteFile{NoStop}{becke2005df}%
\bibitem{adamo1999}%
  \BibitemOpen
  \bibfield{author}{%
  \bibinfo {author} {\bibfnamefont{C.}~\bibnamefont{Adamo}}\ and\ \bibinfo
  {author} {\bibfnamefont{V.}~\bibnamefont{Barone}},\ }%
  \bibfield{journal}{%
  \bibinfo {journal} {J. Chem. Phys.}\ }%
  \textbf{\bibinfo {volume} {110}},\ \bibinfo {pages} {6158} (\bibinfo {year}
  {1999})%
  \bibAnnoteFile{NoStop}{adamo1999}%
\bibitem{tkatchenko2009}%
  \BibitemOpen
  \bibfield{author}{%
  \bibinfo {author} {\bibfnamefont{A.}~\bibnamefont{Tkatchenko}}\ and\ \bibinfo
  {author} {\bibfnamefont{M.}~\bibnamefont{Scheffler}},\ }%
  \bibfield{journal}{%
  \bibinfo {journal} {Phys. Rev. Lett.}\ }%
  \textbf{\bibinfo {volume} {102}},\ \bibinfo {pages} {073005} (\bibinfo {year}
  {2009})%
  \bibAnnoteFile{NoStop}{tkatchenko2009}%
\bibitem{bucko2013}%
  \BibitemOpen
  \bibfield{author}{%
  \bibinfo {author} {\bibfnamefont{T.}~\bibnamefont{Bu\v{c}ko}}, \bibinfo
  {author} {\bibfnamefont{S.}~\bibnamefont{Leb\`{e}gue}}, \bibinfo {author}
  {\bibfnamefont{J.}~\bibnamefont{Hafner}},\ and\ \bibinfo {author}
  {\bibfnamefont{J.~G.}\ \bibnamefont{\'{A}ngy\'{a}n}},\ }%
  \bibfield{journal}{%
  \bibinfo {journal} {J. Chem. Theo. Comput.}\ }%
  \textbf{\bibinfo {volume} {9}},\ \bibinfo {pages} {4293} (\bibinfo {year}
  {2013})%
  \bibAnnoteFile{NoStop}{bucko2013}%
\bibitem{subs}%
  \BibitemOpen
  \enquote{\bibinfo {title} {{\rm These data were obtained on a subset of seven
  crystals with rectangular unit cells.}}.}\ %
  \bibAnnoteFile{NoStop}{subs}%
\bibitem{blochl1994}%
  \BibitemOpen
  \bibfield{author}{%
  \bibinfo {author} {\bibfnamefont{P.~E.}\ \bibnamefont{Bl\"ochl}},\ }%
  \bibfield{journal}{%
  \bibinfo {journal} {Phys. Rev. B}\ }%
  \textbf{\bibinfo {volume} {50}},\ \bibinfo {pages} {17953} (\bibinfo {year}
  {1994})%
  \bibAnnoteFile{NoStop}{blochl1994}%
\bibitem{kresse1999}%
  \BibitemOpen
  \bibfield{author}{%
  \bibinfo {author} {\bibfnamefont{G.}~\bibnamefont{Kresse}}\ and\ \bibinfo
  {author} {\bibfnamefont{J.}~\bibnamefont{Joubert}},\ }%
  \bibfield{journal}{%
  \bibinfo {journal} {Phys. Rev. B}\ }%
  \textbf{\bibinfo {volume} {59}},\ \bibinfo {pages} {1758} (\bibinfo {year}
  {1999})%
  \bibAnnoteFile{NoStop}{kresse1999}%
\bibitem{kaltak2014rpa2}%
  \BibitemOpen
  \bibfield{author}{%
  \bibinfo {author} {\bibfnamefont{M.}~\bibnamefont{Kaltak}}, \bibinfo {author}
  {\bibfnamefont{J.}~\bibnamefont{Klime\v{s}}},\ and\ \bibinfo {author}
  {\bibfnamefont{G.}~\bibnamefont{Kresse}},\ }%
  \bibfield{journal}{%
  \bibinfo {journal} {Phys. Rev. B}\ }%
  \textbf{\bibinfo {volume} {90}},\ \bibinfo {pages} {{054115}} (\bibinfo
  {year} {2014})%
  \bibAnnoteFile{NoStop}{kaltak2014rpa2}%
\bibitem{kaltak2014rpa1}%
  \BibitemOpen
  \bibfield{author}{%
  \bibinfo {author} {\bibfnamefont{M.}~\bibnamefont{Kaltak}}, \bibinfo {author}
  {\bibfnamefont{J.}~\bibnamefont{Klime\v{s}}},\ and\ \bibinfo {author}
  {\bibfnamefont{G.}~\bibnamefont{Kresse}},\ }%
  \bibfield{journal}{%
  \bibinfo {journal} {J. Chem. Theo. Comput.}\ }%
  \textbf{\bibinfo {volume} {10}},\ \bibinfo {pages} {{2498}} (\bibinfo {year}
  {2014})%
  \bibAnnoteFile{NoStop}{kaltak2014rpa1}%
\bibitem{perdew1996}%
  \BibitemOpen
  \bibfield{author}{%
  \bibinfo {author} {\bibfnamefont{J.~P.}\ \bibnamefont{Perdew}}, \bibinfo
  {author} {\bibfnamefont{K.}~\bibnamefont{Burke}},\ and\ \bibinfo {author}
  {\bibfnamefont{M.}~\bibnamefont{Ernzerhof}},\ }%
  \bibfield{journal}{%
  \bibinfo {journal} {Phys. Rev. Lett.}\ }%
  \textbf{\bibinfo {volume} {77}},\ \bibinfo {pages} {3865} (\bibinfo {year}
  {1996}),\ \bibinfo {note} {$ibid$, {\bf 78}, 1396 (1997)}%
  \bibAnnoteFile{NoStop}{perdew1996}%
\bibitem{rozzi2006}%
  \BibitemOpen
  \bibfield{author}{%
  \bibinfo {author} {\bibfnamefont{C.~A.}\ \bibnamefont{Rozzi}}, \bibinfo
  {author} {\bibfnamefont{D.}~\bibnamefont{Varsano}}, \bibinfo {author}
  {\bibfnamefont{A.}~\bibnamefont{Marini}}, \bibinfo {author}
  {\bibfnamefont{E.~K.~U.}\ \bibnamefont{Gross}},\ and\ \bibinfo {author}
  {\bibfnamefont{A.}~\bibnamefont{Rubio}},\ }%
  \bibfield{journal}{%
  \bibinfo {journal} {Phys. Rev. B}\ }%
  \textbf{\bibinfo {volume} {73}},\ \bibinfo {pages} {205119} (\bibinfo {year}
  {2006})%
  \bibAnnoteFile{NoStop}{rozzi2006}%
\bibitem{harl2008}%
  \BibitemOpen
  \bibfield{author}{%
  \bibinfo {author} {\bibfnamefont{J.}~\bibnamefont{Harl}}\ and\ \bibinfo
  {author} {\bibfnamefont{G.}~\bibnamefont{Kresse}},\ }%
  \bibfield{journal}{%
  \bibinfo {journal} {Phys. Rev. B}\ }%
  \textbf{\bibinfo {volume} {77}},\ \bibinfo {pages} {045136} (\bibinfo {year}
  {2008})%
  \bibAnnoteFile{NoStop}{harl2008}%
\bibitem{klimes2014NC}%
  \BibitemOpen
  \bibfield{author}{%
  \bibinfo {author} {\bibfnamefont{J.}~\bibnamefont{Klime\v{s}}}, \bibinfo
  {author} {\bibfnamefont{M.}~\bibnamefont{Kaltak}},\ and\ \bibinfo {author}
  {\bibfnamefont{G.}~\bibnamefont{Kresse}},\ }%
  \bibfield{journal}{%
  \bibinfo {journal} {Phys. Rev. B}\ }%
  \textbf{\bibinfo {volume} {90}},\ \bibinfo {pages} {{075125}} (\bibinfo
  {year} {2014})%
  \bibAnnoteFile{NoStop}{klimes2014NC}%
\bibitem{gulans_unp}%
  \BibitemOpen
  \bibfield{author}{%
  \bibinfo {author} {\bibfnamefont{A.}~\bibnamefont{Gulans}},\ }%
  \bibfield{journal}{%
  \bibinfo {journal} {J. Chem. Phys.}\ }%
  \textbf{\bibinfo {volume} {141}},\ \bibinfo {pages} {{164127}} (\bibinfo
  {year} {2014})%
  \bibAnnoteFile{NoStop}{gulans_unp}%
\bibitem{macher2014}%
  \BibitemOpen
  \bibfield{author}{%
  \bibinfo {author} {\bibfnamefont{M.}~\bibnamefont{Macher}}, \bibinfo {author}
  {\bibfnamefont{J.}~\bibnamefont{Klime\v{s}}}, \bibinfo {author}
  {\bibfnamefont{C.}~\bibnamefont{Franchini}},\ and\ \bibinfo {author}
  {\bibfnamefont{G.}~\bibnamefont{Kresse}},\ }%
  \bibfield{journal}{%
  \bibinfo {journal} {J. Chem. Phys.}\ }%
  \textbf{\bibinfo {volume} {140}},\ \bibinfo {pages} {{084502}} (\bibinfo
  {year} {2014})%
  \bibAnnoteFile{NoStop}{macher2014}%
\bibitem{lu2009}%
  \BibitemOpen
  \bibfield{author}{%
  \bibinfo {author} {\bibfnamefont{D.}~\bibnamefont{Lu}}, \bibinfo {author}
  {\bibfnamefont{Y.}~\bibnamefont{Li}}, \bibinfo {author}
  {\bibfnamefont{D.}~\bibnamefont{Rocca}},\ and\ \bibinfo {author}
  {\bibfnamefont{G.}~\bibnamefont{Galli}},\ }%
  \bibfield{journal}{%
  \bibinfo {journal} {Phys. Rev. Lett.}\ }%
  \textbf{\bibinfo {volume} {102}},\ \bibinfo {pages} {206411} (\bibinfo {year}
  {2009})%
  \bibAnnoteFile{NoStop}{lu2009}%
\bibitem{ambrosetti2014}%
  \BibitemOpen
  \bibfield{author}{%
  \bibinfo {author} {\bibfnamefont{A.}~\bibnamefont{Ambrosetti}}, \bibinfo
  {author} {\bibfnamefont{A.~M.}\ \bibnamefont{Reilly}}, \bibinfo {author}
  {\bibfnamefont{R.~A.}\ \bibnamefont{DiStasio}},\ and\ \bibinfo {author}
  {\bibfnamefont{A.}~\bibnamefont{Tkatchenko}},\ }%
  \bibfield{journal}{%
  \bibinfo {journal} {J. Chem. Phys.}\ }%
  \textbf{\bibinfo {volume} {140}},\ \bibinfo {pages} {18A508} (\bibinfo {year}
  {2014})%
  \bibAnnoteFile{NoStop}{ambrosetti2014}%
\bibitem{paier2005}%
  \BibitemOpen
  \bibfield{author}{%
  \bibinfo {author} {\bibfnamefont{J.}~\bibnamefont{Paier}}, \bibinfo {author}
  {\bibfnamefont{R.}~\bibnamefont{Hirschl}}, \bibinfo {author}
  {\bibfnamefont{M.}~\bibnamefont{Marsman}},\ and\ \bibinfo {author}
  {\bibfnamefont{G.}~\bibnamefont{Kresse}},\ }%
  \bibfield{journal}{%
  \bibinfo {journal} {J. Chem. Phys.}\ }%
  \textbf{\bibinfo {volume} {122}},\ \bibinfo {pages} {234102} (\bibinfo {year}
  {2005})%
  \bibAnnoteFile{NoStop}{paier2005}%
\bibitem{spencer2008}%
  \BibitemOpen
  \bibfield{author}{%
  \bibinfo {author} {\bibfnamefont{J.}~\bibnamefont{Spencer}}\ and\ \bibinfo
  {author} {\bibfnamefont{A.}~\bibnamefont{Alavi}},\ }%
  \bibfield{journal}{%
  \bibinfo {journal} {Phys. Rev. B}\ }%
  \textbf{\bibinfo {volume} {77}},\ \bibinfo {pages} {193110} (\bibinfo {year}
  {2008})%
  \bibAnnoteFile{NoStop}{spencer2008}%
\bibitem{beran2016}%
  \BibitemOpen
  \bibfield{author}{%
  \bibinfo {author} {\bibfnamefont{G.~J.~O.}\ \bibnamefont{Beran}},\ }%
  \bibfield{journal}{%
  \bibinfo {journal} {Chem. Rev.}\ }%
  \textbf{\bibinfo {volume} {116}},\ \bibinfo {pages} {5567} (\bibinfo {year}
  {2016})%
  \bibAnnoteFile{NoStop}{beran2016}%
\bibitem{moellmann2014}%
  \BibitemOpen
  \bibfield{author}{%
  \bibinfo {author} {\bibfnamefont{J.}~\bibnamefont{Moellmann}}\ and\ \bibinfo
  {author} {\bibfnamefont{S.}~\bibnamefont{Grimme}},\ }%
  \bibfield{journal}{%
  \bibinfo {journal} {J. Phys. Chem. C}\ }%
  \textbf{\bibinfo {volume} {118}},\ \bibinfo {pages} {7615} (\bibinfo {year}
  {2014})%
  \bibAnnoteFile{NoStop}{moellmann2014}%
\bibitem{g09}%
  \BibitemOpen
  \bibfield{author}{%
  \bibinfo {author} {\bibfnamefont{M.~J.}\ \bibnamefont{Frisch}}, \bibinfo
  {author} {\bibfnamefont{G.~W.}\ \bibnamefont{Trucks}}, \bibinfo {author}
  {\bibfnamefont{H.~B.}\ \bibnamefont{Schlegel}}, \bibinfo {author}
  {\bibfnamefont{G.~E.}\ \bibnamefont{Scuseria}}, \bibinfo {author}
  {\bibfnamefont{M.~A.}\ \bibnamefont{Robb}}, \bibinfo {author}
  {\bibfnamefont{J.~R.}\ \bibnamefont{Cheeseman}}, \bibinfo {author}
  {\bibfnamefont{G.}~\bibnamefont{Scalmani}}, \bibinfo {author}
  {\bibfnamefont{V.}~\bibnamefont{Barone}}, \bibinfo {author}
  {\bibfnamefont{B.}~\bibnamefont{Mennucci}}, \bibinfo {author}
  {\bibfnamefont{G.~A.}\ \bibnamefont{Petersson}}, \bibinfo {author}
  {\bibfnamefont{H.}~\bibnamefont{Nakatsuji}}, \bibinfo {author}
  {\bibfnamefont{M.}~\bibnamefont{Caricato}}, \bibinfo {author}
  {\bibfnamefont{X.}~\bibnamefont{Li}}, \bibinfo {author}
  {\bibfnamefont{H.~P.}\ \bibnamefont{Hratchian}}, \bibinfo {author}
  {\bibfnamefont{A.~F.}\ \bibnamefont{Izmaylov}}, \bibinfo {author}
  {\bibfnamefont{J.}~\bibnamefont{Bloino}}, \bibinfo {author}
  {\bibfnamefont{G.}~\bibnamefont{Zheng}}, \bibinfo {author}
  {\bibfnamefont{J.~L.}\ \bibnamefont{Sonnenberg}}, \bibinfo {author}
  {\bibfnamefont{M.}~\bibnamefont{Hada}}, \bibinfo {author}
  {\bibfnamefont{M.}~\bibnamefont{Ehara}}, \bibinfo {author}
  {\bibfnamefont{K.}~\bibnamefont{Toyota}}, \bibinfo {author}
  {\bibfnamefont{R.}~\bibnamefont{Fukuda}}, \bibinfo {author}
  {\bibfnamefont{J.}~\bibnamefont{Hasegawa}}, \bibinfo {author}
  {\bibfnamefont{M.}~\bibnamefont{Ishida}}, \bibinfo {author}
  {\bibfnamefont{T.}~\bibnamefont{Nakajima}}, \bibinfo {author}
  {\bibfnamefont{Y.}~\bibnamefont{Honda}}, \bibinfo {author}
  {\bibfnamefont{O.}~\bibnamefont{Kitao}}, \bibinfo {author}
  {\bibfnamefont{H.}~\bibnamefont{Nakai}}, \bibinfo {author}
  {\bibfnamefont{T.}~\bibnamefont{Vreven}}, \bibinfo {author}
  {\bibfnamefont{J.~A.}\ \bibnamefont{Montgomery}, \bibfnamefont{{Jr.}}},
  \bibinfo {author} {\bibfnamefont{J.~E.}\ \bibnamefont{Peralta}}, \bibinfo
  {author} {\bibfnamefont{F.}~\bibnamefont{Ogliaro}}, \bibinfo {author}
  {\bibfnamefont{M.}~\bibnamefont{Bearpark}}, \bibinfo {author}
  {\bibfnamefont{J.~J.}\ \bibnamefont{Heyd}}, \bibinfo {author}
  {\bibfnamefont{E.}~\bibnamefont{Brothers}}, \bibinfo {author}
  {\bibfnamefont{K.~N.}\ \bibnamefont{Kudin}}, \bibinfo {author}
  {\bibfnamefont{V.~N.}\ \bibnamefont{Staroverov}}, \bibinfo {author}
  {\bibfnamefont{R.}~\bibnamefont{Kobayashi}}, \bibinfo {author}
  {\bibfnamefont{J.}~\bibnamefont{Normand}}, \bibinfo {author}
  {\bibfnamefont{K.}~\bibnamefont{Raghavachari}}, \bibinfo {author}
  {\bibfnamefont{A.}~\bibnamefont{Rendell}}, \bibinfo {author}
  {\bibfnamefont{J.~C.}\ \bibnamefont{Burant}}, \bibinfo {author}
  {\bibfnamefont{S.~S.}\ \bibnamefont{Iyengar}}, \bibinfo {author}
  {\bibfnamefont{J.}~\bibnamefont{Tomasi}}, \bibinfo {author}
  {\bibfnamefont{M.}~\bibnamefont{Cossi}}, \bibinfo {author}
  {\bibfnamefont{N.}~\bibnamefont{Rega}}, \bibinfo {author}
  {\bibfnamefont{J.~M.}\ \bibnamefont{Millam}}, \bibinfo {author}
  {\bibfnamefont{M.}~\bibnamefont{Klene}}, \bibinfo {author}
  {\bibfnamefont{J.~E.}\ \bibnamefont{Knox}}, \bibinfo {author}
  {\bibfnamefont{J.~B.}\ \bibnamefont{Cross}}, \bibinfo {author}
  {\bibfnamefont{V.}~\bibnamefont{Bakken}}, \bibinfo {author}
  {\bibfnamefont{C.}~\bibnamefont{Adamo}}, \bibinfo {author}
  {\bibfnamefont{J.}~\bibnamefont{Jaramillo}}, \bibinfo {author}
  {\bibfnamefont{R.}~\bibnamefont{Gomperts}}, \bibinfo {author}
  {\bibfnamefont{R.~E.}\ \bibnamefont{Stratmann}}, \bibinfo {author}
  {\bibfnamefont{O.}~\bibnamefont{Yazyev}}, \bibinfo {author}
  {\bibfnamefont{A.~J.}\ \bibnamefont{Austin}}, \bibinfo {author}
  {\bibfnamefont{R.}~\bibnamefont{Cammi}}, \bibinfo {author}
  {\bibfnamefont{C.}~\bibnamefont{Pomelli}}, \bibinfo {author}
  {\bibfnamefont{J.~W.}\ \bibnamefont{Ochterski}}, \bibinfo {author}
  {\bibfnamefont{R.~L.}\ \bibnamefont{Martin}}, \bibinfo {author}
  {\bibfnamefont{K.}~\bibnamefont{Morokuma}}, \bibinfo {author}
  {\bibfnamefont{V.~G.}\ \bibnamefont{Zakrzewski}}, \bibinfo {author}
  {\bibfnamefont{G.~A.}\ \bibnamefont{Voth}}, \bibinfo {author}
  {\bibfnamefont{P.}~\bibnamefont{Salvador}}, \bibinfo {author}
  {\bibfnamefont{J.~J.}\ \bibnamefont{Dannenberg}}, \bibinfo {author}
  {\bibfnamefont{S.}~\bibnamefont{Dapprich}}, \bibinfo {author}
  {\bibfnamefont{A.~D.}\ \bibnamefont{Daniels}}, \bibinfo {author}
  {\bibfnamefont{O.}~\bibnamefont{Farkas}}, \bibinfo {author}
  {\bibfnamefont{J.~B.}\ \bibnamefont{Foresman}}, \bibinfo {author}
  {\bibfnamefont{J.~V.}\ \bibnamefont{Ortiz}}, \bibinfo {author}
  {\bibfnamefont{J.}~\bibnamefont{Cioslowski}},\ and\ \bibinfo {author}
  {\bibfnamefont{D.~J.}\ \bibnamefont{Fox}},\ }%
  \enquote{\bibinfo {title} {Gaussian∼09 {R}evision {E}.01},}\ \bibinfo
  {note} {Gaussian Inc. Wallingford CT 2009}%
  \bibAnnoteFile{NoStop}{g09}%
\bibitem{ruiz2012}%
  \BibitemOpen
  \bibfield{author}{%
  \bibinfo {author} {\bibfnamefont{V.~G.}\ \bibnamefont{Ruiz}}, \bibinfo
  {author} {\bibfnamefont{W.}~\bibnamefont{Liu}}, \bibinfo {author}
  {\bibfnamefont{E.}~\bibnamefont{Zojer}}, \bibinfo {author}
  {\bibfnamefont{M.}~\bibnamefont{Scheffler}},\ and\ \bibinfo {author}
  {\bibfnamefont{A.}~\bibnamefont{Tkatchenko}},\ }%
  \bibfield{journal}{%
  \bibinfo {journal} {Phys. Rev. Lett.}\ }%
  \textbf{\bibinfo {volume} {108}},\ \bibinfo {pages} {146103} (\bibinfo {year}
  {2012})%
  \bibAnnoteFile{NoStop}{ruiz2012}%
\bibitem{bleiziffer2013}%
  \BibitemOpen
  \bibfield{author}{%
  \bibinfo {author} {\bibfnamefont{P.}~\bibnamefont{Bleiziffer}}, \bibinfo
  {author} {\bibfnamefont{A.}~\bibnamefont{He{\ss}elmann}},\ and\ \bibinfo
  {author} {\bibfnamefont{A.}~\bibnamefont{G{\"o}rling}},\ }%
  \bibfield{journal}{%
  \bibinfo {journal} {J. Chem. Phys.}\ }%
  \textbf{\bibinfo {volume} {{139}}},\ \bibinfo {pages} {{084113}} (\bibinfo
  {year} {{2013}})%
  \bibAnnoteFile{NoStop}{bleiziffer2013}%
\bibitem{klimes2014}%
  \BibitemOpen
  \bibfield{author}{%
  \bibinfo {author} {\bibfnamefont{J.}~\bibnamefont{Klime\v{s}}}\ and\ \bibinfo
  {author} {\bibfnamefont{G.}~\bibnamefont{Kresse}},\ }%
  \bibfield{journal}{%
  \bibinfo {journal} {J. Chem. Phys.}\ }%
  \textbf{\bibinfo {volume} {140}},\ \bibinfo {pages} {{054516}} (\bibinfo
  {year} {2014})%
  \bibAnnoteFile{NoStop}{klimes2014}%
\bibitem{grueneis2009}%
  \BibitemOpen
  \bibfield{author}{%
  \bibinfo {author} {\bibfnamefont{A.}~\bibnamefont{Gr\"uneis}}, \bibinfo
  {author} {\bibfnamefont{M.}~\bibnamefont{Marsman}}, \bibinfo {author}
  {\bibfnamefont{J.}~\bibnamefont{Harl}}, \bibinfo {author}
  {\bibfnamefont{L.}~\bibnamefont{Schimka}},\ and\ \bibinfo {author}
  {\bibfnamefont{G.}~\bibnamefont{Kresse}},\ }%
  \bibfield{journal}{%
  \bibinfo {journal} {J. Chem. Phys.}\ }%
  \textbf{\bibinfo {volume} {{131}}},\ \bibinfo {pages} {{154115}} (\bibinfo
  {year} {{2009}})%
  \bibAnnoteFile{NoStop}{grueneis2009}%
\bibitem{paier2012}%
  \BibitemOpen
  \bibfield{author}{%
  \bibinfo {author} {\bibfnamefont{J.}~\bibnamefont{Paier}}, \bibinfo {author}
  {\bibfnamefont{X.}~\bibnamefont{Ren}}, \bibinfo {author}
  {\bibfnamefont{P.}~\bibnamefont{Rinke}}, \bibinfo {author}
  {\bibfnamefont{G.~E.}\ \bibnamefont{Scuseria}}, \bibinfo {author}
  {\bibfnamefont{A.}~\bibnamefont{Gr\"{u}neis}}, \bibinfo {author}
  {\bibfnamefont{G.}~\bibnamefont{Kresse}},\ and\ \bibinfo {author}
  {\bibfnamefont{M.}~\bibnamefont{Scheffler}},\ }%
  \bibfield{journal}{%
  \bibinfo {journal} {New J. Phys.}\ }%
  \textbf{\bibinfo {volume} {14}},\ \bibinfo {pages} {043002} (\bibinfo {year}
  {2012})%
  \bibAnnoteFile{NoStop}{paier2012}%
\bibitem{olsen2013}%
  \BibitemOpen
  \bibfield{author}{%
  \bibinfo {author} {\bibfnamefont{T.}~\bibnamefont{Olsen}}\ and\ \bibinfo
  {author} {\bibfnamefont{K.~S.}\ \bibnamefont{Thygesen}},\ }%
  \bibfield{journal}{%
  \bibinfo {journal} {Phys. Rev. B}\ }%
  \textbf{\bibinfo {volume} {88}},\ \bibinfo {pages} {115131} (\bibinfo {year}
  {2013})%
  \bibAnnoteFile{NoStop}{olsen2013}%
\bibitem{olsen2014}%
  \BibitemOpen
  \bibfield{author}{%
  \bibinfo {author} {\bibfnamefont{T.}~\bibnamefont{Olsen}}\ and\ \bibinfo
  {author} {\bibfnamefont{K.~S.}\ \bibnamefont{Thygesen}},\ }%
  \bibfield{journal}{%
  \bibinfo {journal} {Phys. Rev. Lett.}\ }%
  \textbf{\bibinfo {volume} {112}},\ \bibinfo {pages} {203001} (\bibinfo {year}
  {2014})%
  \bibAnnoteFile{NoStop}{olsen2014}%
\end{thebibliography}

%

\end{document}